\documentclass[acmsmall,sigplan,]{acmart}
\acmConference[IFL2020]{32nd International Symposium on Implementation and Application of Functional Languages}{September, 2020}{}
\acmYear{2020}
\acmISBN{} % \acmISBN{978-x-xxxx-xxxx-x/YY/MM}
\acmDOI{} % \acmDOI{10.1145/nnnnnnn.nnnnnnn}
\startPage{1}

\setcopyright{none}
\usepackage{xurl}
\usepackage{booktabs}   %% For formal tables:
\usepackage{subcaption} %% For complex figures with subfigures/subcaptions
\usepackage{tikz-cd}

\usepackage[utf8x]{inputenc}
\usepackage{newunicodechar}
\newunicodechar{∀}{\ensuremath{\forall{}}}
\providecommand{\tightlist}{%
  \setlength{\itemsep}{0pt}\setlength{\parskip}{0pt}}

\newenvironment{Shaded}{}{}
\usepackage{fancyvrb}
\newcommand{\AlertTok}[1]{\textcolor[rgb]{1.00,0.00,0.00}{#1}}
\newcommand{\AnnotationTok}[1]{\textcolor[rgb]{0.00,0.50,0.00}{#1}}
\newcommand{\AttributeTok}[1]{#1}
\newcommand{\BaseNTok}[1]{#1}
\newcommand{\BuiltInTok}[1]{#1}
\newcommand{\CharTok}[1]{\textcolor[rgb]{0.00,0.50,0.50}{#1}}
\newcommand{\CommentTok}[1]{\textcolor[rgb]{0.00,0.50,0.00}{#1}}
\newcommand{\CommentVarTok}[1]{\textcolor[rgb]{0.00,0.50,0.00}{#1}}
\newcommand{\ConstantTok}[1]{#1}
\newcommand{\ControlFlowTok}[1]{\textcolor[rgb]{0.00,0.00,1.00}{#1}}
\newcommand{\DataTypeTok}[1]{#1}
\newcommand{\DecValTok}[1]{#1}
\newcommand{\DocumentationTok}[1]{\textcolor[rgb]{0.00,0.50,0.00}{#1}}
\newcommand{\ErrorTok}[1]{\textcolor[rgb]{1.00,0.00,0.00}{\textbf{#1}}}
\newcommand{\ExtensionTok}[1]{#1}
\newcommand{\FloatTok}[1]{#1}
\newcommand{\FunctionTok}[1]{#1}
\newcommand{\ImportTok}[1]{#1}
\newcommand{\InformationTok}[1]{\textcolor[rgb]{0.00,0.50,0.00}{#1}}
\newcommand{\KeywordTok}[1]{\textcolor[rgb]{0.00,0.00,1.00}{#1}}
\newcommand{\NormalTok}[1]{#1}
\newcommand{\OperatorTok}[1]{#1}
\newcommand{\OtherTok}[1]{\textcolor[rgb]{1.00,0.25,0.00}{#1}}
\newcommand{\PreprocessorTok}[1]{\textcolor[rgb]{1.00,0.25,0.00}{#1}}
\newcommand{\RegionMarkerTok}[1]{#1}
\newcommand{\SpecialCharTok}[1]{\textcolor[rgb]{0.00,0.50,0.50}{#1}}
\newcommand{\SpecialStringTok}[1]{\textcolor[rgb]{0.00,0.50,0.50}{#1}}
\newcommand{\StringTok}[1]{\textcolor[rgb]{0.00,0.50,0.50}{#1}}
\newcommand{\VariableTok}[1]{#1}
\newcommand{\VerbatimStringTok}[1]{\textcolor[rgb]{0.00,0.50,0.50}{#1}}
\newcommand{\WarningTok}[1]{\textcolor[rgb]{0.00,0.50,0.00}{\textbf{#1}}}
\DefineVerbatimEnvironment{Highlighting}{Verbatim}{commandchars=\\\{\}}

\newlength{\cslhangindent}
\newenvironment{cslreferences}%
  {\setlength{\parindent}{0pt}%
  \everypar{\setlength{\hangindent}{\cslhangindent}}\ignorespaces}%
  {\par}

\begin{document}

\title[Perfect union type]{Towards a
more perfect union type}
    \author{Michał J. Gajda}
    \affiliation{
   \
   \
   \
   \
   \
   \
   \
   \
  }
      \date{2020-06-19}
%% Title information
\title{Towards a more perfect union
type}         %% [Short Title] is optional;
                                        %% when present, will be used in
                                        %% header instead of Full Title.
             %% \titlenote is optional;
                                        %% can be repeated if necessary;
                                        %% contents suppressed with 'anonymous'
                     %% \subtitle is optional
       %% \subtitlenote is optional;
                                        %% can be repeated if necessary;
                                        %% contents suppressed with 'anonymous'

%% Author information
%% Contents and number of authors suppressed with 'anonymous'.
%% Each author should be introduced by \author, followed by
%% \authornote (optional), \orcid (optional), \affiliation, and
%% \email.
%% An author may have multiple affiliations and/or emails; repeat the
%% appropriate command.
%% Many elements are not rendered, but should be provided for metadata
%% extraction tools.

%% Abstract
%% Note: \begin{abstract}...\end{abstract} environment must come
%% before \maketitle command
\begin{abstract}
We present a principled theoretical
framework for inferring and checking the
union types, and show its work in
practice on JSON data structures.

The framework poses a union type
inference as a learning problem from
multiple examples. The categorical
framework is generic and easily
extensible.
\end{abstract}

%% 2012 ACM Computing Classification System (CSS) concepts
%% Generate at 'http://dl.acm.org/ccs/ccs.cfm'.

%% End of generated code

%% Keywords
%% comma separated list

%% \maketitle
%% Note: \maketitle command must come after title commands, author
%% commands, abstract environment, Computing Classification System
%% environment and commands, and keywords command.
\maketitle

\hypertarget{introduction}{%
\section{Introduction}\label{introduction}}

Typing dynamic languages has been long
considered a challenge
{[}\protect\hyperlink{ref-javascript-inference}{3}{]}.
The importance of the task grown with
the ubiquity cloud application
programming interfaces (APIs) utilizing
JavaScript object notation (JSON), where
one needs to infer the structure having
only a limited number of sample
documents available. Previous research
has suggested it is possible to infer
adequate type mappings from sample data
{[}\protect\hyperlink{ref-quicktype}{2},
\protect\hyperlink{ref-pads}{8},
\protect\hyperlink{ref-json-autotype-prezi}{14},
\protect\hyperlink{ref-type-providers-f-sharp}{20}{]}.

In the present study, we expand on these
results. We propose a modular framework
for type systems in programming
languages as learning algorithms,
formulate it as equational identities,
and evaluate its performance on
inference of Haskell data types from
JSON API examples.

\hypertarget{related-work}{%
\subsection{Related
work}\label{related-work}}

\hypertarget{union-type-providers}{%
\subsubsection{Union type
providers}\label{union-type-providers}}

The earliest practical effort to apply
union types to JSON inference to
generate Haskell types
{[}\protect\hyperlink{ref-json-autotype-prezi}{14}{]}.
It uses union type theory, but it also
lacks an extensible theoretical
framework. F\# type providers for JSON
facilitate deriving a schema
automatically; however, a type system
does not support union of alternatives
and is given shape inference algorithm,
instead of design driven by desired
properties
{[}\protect\hyperlink{ref-type-providers-f-sharp}{20}{]}.
The other attempt to automatically infer
schemas has been introduced in the PADS
project
{[}\protect\hyperlink{ref-pads}{8}{]}.
Nevertheless, it has not specified a
generalized type-system design
methodology. One approach uses Markov
chains to derive JSON types
{[}\protect\hyperlink{ref-quicktype}{2}{]}\footnote{This
  approach uses Markov chains to infer
  best of alternative type
  representations.}. This approach
requires considerable engineering time
due to the implementation of unit tests
in a case-by-case mode, instead of
formulating laws applying to all types.
Moreover, this approach lacks a sound
underlying theory. Regular expression
types were also used to type XML
documents
{[}\protect\hyperlink{ref-xduce}{13}{]},
which does not allow for selecting
alternative representation. In the
present study, we generalize previously
introduced approaches and enable a
systematic addition of not only value
sets, but inference subalgorithms, to
the union type system.

\hypertarget{frameworks-for-describing-type-systems}{%
\subsubsection{Frameworks for describing
type
systems}\label{frameworks-for-describing-type-systems}}

Type systems are commonly expressed as
partial relation of \emph{typing}. Their
properties, such as subject reduction
are also expressed relatively to the
relation (also partial) of
\emph{reduction} within a term rewriting
system. General formulations have been
introduced for the Damas-Milner type
systems parameterized by constraints
{[}\protect\hyperlink{ref-HM-X}{23}{]}.
It is also worth noting that traditional
Damas-Milner type disciplines enjoy
decidability, and embrace the laws of
soundness, and subject-reduction.
However these laws often prove too
strict during type system extension,
dependent type systems often abandon
subject-reduction, and type systems of
widely used programming languages are
either undecidable
{[}\protect\hyperlink{ref-GHCZurihac}{21}{]},
or even unsound
{[}\protect\hyperlink{ref-typescript-soundness}{27}{]}.

Early approaches used lattice structure
on the types
{[}\protect\hyperlink{ref-subtyping-lattice}{25}{]},
which is more stringent than ours since
it requires idempotence of unification
(as join operation), as well as
complementary meet operation with the
same properties. Semantic subtyping
approach provides a characterization of
a set-based union, intersection, and
complement types
{[}\protect\hyperlink{ref-semantic-subtyping}{9},
\protect\hyperlink{ref-semantic-subtyping2}{10}{]},
which allows model subtype containment
on first-order types and functions. This
model relies on building a model using
infinite sets in set theory, but its
rigidity fails to generalize to
non-idempotent learning\footnote{Which
  would allow extension with machine
  learning techniques like Markov chains
  to infer optimal type representation
  from frequency of occuring
  values{[}\protect\hyperlink{ref-quicktype}{2}{]}.}.
We are also not aware of a type
inference framework that consistently
and completely preserve information in
the face of inconsistencies nor errors,
beyond using \texttt{bottom} and
expanding to \emph{infamous undefined
behaviour}
{[}\protect\hyperlink{ref-undefined1}{5}{]}.

We propose a categorical and
constructive framework that preserves
the soundness in inference while
allowing for consistent approximations.
Indeed our experience is that most of
the type system implementation may be
generic.

\hypertarget{sec:examples}{%
\section{Motivation}\label{sec:examples}}

Here, we consider several examples
similar to JSON API descriptions. We
provide these examples in the form of a
few JSON objects, along with desired
representation as Haskell data
declaration.

\begin{enumerate}
\def\labelenumi{\arabic{enumi}.}
\item
  Subsets of data within a single
  constructor:

  \begin{enumerate}
  \def\labelenumii{\alph{enumii}.}
  \tightlist
  \item
    \emph{API argument is an email} --
    it is a subset of valid
    \texttt{String} values that can be
    validated on the client-side.
  \item
    \emph{The page size determines the
    number of results to return (min:
    10, max:10,000)} -- it is also a
    subset of integer values
    (\texttt{Int}) between \(10\), and
    \(10,000\)
  \item
    \emph{The \texttt{date} field
    contains ISO8601 date} -- a record
    field represented as a
    \texttt{String} that contains a
    calendar date in the format
    \texttt{"2019-03-03"}
  \end{enumerate}
\end{enumerate}

\begin{enumerate}
\def\labelenumi{\arabic{enumi}.}
\setcounter{enumi}{1}
\tightlist
\item
  Optional fields: \emph{The page size
  is equal to 100 by default} -- it
  means we expect to see the record like
  \texttt{\{"page\_size":\ 50\}} or an
  empty record \texttt{\{\}} that should
  be interpreted in the same way as
  \texttt{\{"page\_size":\ 100\}}
\end{enumerate}

\begin{enumerate}
\def\labelenumi{\arabic{enumi}.}
\setcounter{enumi}{2}
\tightlist
\item
  Variant fields: \emph{Answer to a
  query is either a number of registered
  objects, or String
  \texttt{"unavailable"}} - this is
  integer value (\texttt{Int}) or a
  \texttt{String}
  (\texttt{Int\ :\textbar{}:\ String})
\end{enumerate}

\begin{enumerate}
\def\labelenumi{\arabic{enumi}.}
\setcounter{enumi}{3}
\tightlist
\item
  Variant records: \emph{Answer contains
  either a text message with a user
  identifier or an error.} -- That can
  be represented as one of following
  options:
\end{enumerate}

\begin{Shaded}
\begin{Highlighting}[]
\FunctionTok{\{}\DataTypeTok{"message"}\FunctionTok{:} \StringTok{"Where can I submit proposal?"}\FunctionTok{,} \DataTypeTok{"uid"} \FunctionTok{:} \DecValTok{1014}\FunctionTok{\}}
\FunctionTok{\{}\DataTypeTok{"message"}\FunctionTok{:} \StringTok{"Submit it to HotCRP"}\FunctionTok{,}          \DataTypeTok{"uid"} \FunctionTok{:}  \DecValTok{317}\FunctionTok{\}}
\FunctionTok{\{}\DataTypeTok{"error"}  \FunctionTok{:} \StringTok{"Authorization failed"}\FunctionTok{,}         \DataTypeTok{"code"}\FunctionTok{:}  \DecValTok{401}\FunctionTok{\}}
\FunctionTok{\{}\DataTypeTok{"error"}  \FunctionTok{:} \StringTok{"User not found"}\FunctionTok{,}               \DataTypeTok{"code"}\FunctionTok{:}  \DecValTok{404}\FunctionTok{\}}
\end{Highlighting}
\end{Shaded}

\setlength{\tabcolsep}{1pt}
\begin{tabular}{lllllllllllllllllll}
\multicolumn{3}{l}{$\textbf{data}\textit{ }\textsc{Example4}\textit{ }\textit{=}\textit{ }\textsc{Message}\textit{ }$} & \multicolumn{1}{l}{$\textit{\{}$} & \multicolumn{1}{l}{$\textit{ }$} & \multicolumn{1}{l}{$\emph{message}\textit{ }$} & \multicolumn{1}{l}{$\textit{::}$} & \multicolumn{1}{l}{$\textit{ }$} & \multicolumn{1}{l}{$\textsc{String}\textit{
}$}\\
\multicolumn{3}{p{24ex}}{                        } & \multicolumn{1}{l}{$\textit{,}$} & \multicolumn{1}{l}{$\textit{ }$} & \multicolumn{1}{l}{$\emph{uid}\textit{     }$} & \multicolumn{1}{l}{$\textit{::}$} & \multicolumn{1}{l}{$\textit{ }$} & \multicolumn{1}{l}{$\textsc{Int}\textit{ }\textit{\}}\textit{
}$}\\
\multicolumn{2}{p{14ex}}{              } & \multicolumn{1}{l}{$\textit{|}\textit{ }\textsc{Error}\textit{   }$} & \multicolumn{1}{l}{$\textit{\{}$} & \multicolumn{1}{l}{$\textit{ }$} & \multicolumn{1}{l}{$\emph{error}\textit{   }$} & \multicolumn{1}{l}{$\textit{::}$} & \multicolumn{1}{l}{$\textit{ }$} & \multicolumn{1}{l}{$\textsc{String}\textit{
}$}\\
\multicolumn{3}{p{24ex}}{                        } & \multicolumn{1}{l}{$\textit{,}$} & \multicolumn{1}{l}{$\textit{ }$} & \multicolumn{1}{l}{$\emph{code}\textit{    }$} & \multicolumn{1}{l}{$\textit{::}$} & \multicolumn{1}{l}{$\textit{ }$} & \multicolumn{1}{l}{$\textsc{Int}\textit{ }\textit{\}}$}\\

\end{tabular}

\begin{enumerate}
\def\labelenumi{\arabic{enumi}.}
\setcounter{enumi}{4}
\tightlist
\item
  Arrays corresponding to records:
\end{enumerate}

\hypertarget{lst:row-constraint}{%
\label{lst:row-constraint}}%
\begin{Shaded}
\begin{Highlighting}[]
\OtherTok{[} \OtherTok{[}\DecValTok{1}\OtherTok{,} \StringTok{"Nick"}\OtherTok{,}    \KeywordTok{null}       \OtherTok{]}
\OtherTok{,} \OtherTok{[}\DecValTok{2}\OtherTok{,} \StringTok{"George"}\OtherTok{,} \StringTok{"2019{-}04{-}11"}\OtherTok{]}
\OtherTok{,} \OtherTok{[}\DecValTok{3}\OtherTok{,} \StringTok{"Olivia"}\OtherTok{,} \StringTok{"1984{-}05{-}03"}\OtherTok{]} \OtherTok{]}
\end{Highlighting}
\end{Shaded}

\begin{enumerate}
\def\labelenumi{\arabic{enumi}.}
\setcounter{enumi}{5}
\tightlist
\item
  Maps of identical objects (example
  from
  {[}\protect\hyperlink{ref-quicktype}{2}{]}):
\end{enumerate}

\begin{Shaded}
\begin{Highlighting}[]
\FunctionTok{\{}   \DataTypeTok{"6408f5"}\FunctionTok{:} \FunctionTok{\{} \DataTypeTok{"size"}\FunctionTok{:}       \DecValTok{969709}    \FunctionTok{,} \DataTypeTok{"height"}\FunctionTok{:}    \DecValTok{510599}
              \FunctionTok{,} \DataTypeTok{"difficulty"}\FunctionTok{:} \FloatTok{866429.732}\FunctionTok{,} \DataTypeTok{"previous"}\FunctionTok{:} \StringTok{"54fced"} \FunctionTok{\},}
    \DataTypeTok{"54fced"}\FunctionTok{:} \FunctionTok{\{} \DataTypeTok{"size"}\FunctionTok{:}       \DecValTok{991394}    \FunctionTok{,} \DataTypeTok{"height"}\FunctionTok{:}    \DecValTok{510598}
              \FunctionTok{,} \DataTypeTok{"difficulty"}\FunctionTok{:} \FloatTok{866429.823}\FunctionTok{,} \DataTypeTok{"previous"}\FunctionTok{:} \StringTok{"6c9589"} \FunctionTok{\},}
    \DataTypeTok{"6c9589"}\FunctionTok{:} \FunctionTok{\{} \DataTypeTok{"size"}\FunctionTok{:}       \DecValTok{990527}    \FunctionTok{,} \DataTypeTok{"height"}\FunctionTok{:}    \DecValTok{510597}
              \FunctionTok{,} \DataTypeTok{"difficulty"}\FunctionTok{:} \FloatTok{866429.931}\FunctionTok{,} \DataTypeTok{"previous"}\FunctionTok{:} \StringTok{"51a0cb"} \FunctionTok{\}} \FunctionTok{\}}
\end{Highlighting}
\end{Shaded}

It should be noted that the last example
presented above requires Haskell
representation inference to be
non-monotonic, as an example of object
with only a single key would be best
represented by a record type:

\setlength{\tabcolsep}{1pt}
\begin{tabular}{lllllllllllllllllllllllllll}
\multicolumn{2}{l}{$\textbf{data}\textit{ }\textsc{Example}\textit{ }\textit{=}\textit{ }\textsc{Example}\textit{ }$} & \multicolumn{1}{l}{$\textit{\{}$} & \multicolumn{1}{l}{$\textit{ }$} & \multicolumn{3}{l}{$\emph{f\_{}6408f5}$} & \multicolumn{1}{l}{$\textit{ }$} & \multicolumn{1}{l}{$\textit{::}$} & \multicolumn{1}{l}{$\textit{ }$} & \multicolumn{4}{l}{$\textsc{O\_{}6408f5}$} & \multicolumn{3}{l}{$\textit{
}$}\\
\multicolumn{2}{p{23ex}}{                       } & \multicolumn{1}{l}{$\textit{,}$} & \multicolumn{1}{l}{$\textit{ }$} & \multicolumn{3}{l}{$\emph{f\_{}54fced}$} & \multicolumn{1}{l}{$\textit{ }$} & \multicolumn{1}{l}{$\textit{::}$} & \multicolumn{1}{l}{$\textit{ }$} & \multicolumn{4}{l}{$\textsc{O\_{}6408f5}$} & \multicolumn{3}{l}{$\textit{
}$}\\
\multicolumn{2}{p{23ex}}{                       } & \multicolumn{1}{l}{$\textit{,}$} & \multicolumn{1}{l}{$\textit{ }$} & \multicolumn{3}{l}{$\emph{f\_{}6c9589}$} & \multicolumn{1}{l}{$\textit{ }$} & \multicolumn{1}{l}{$\textit{::}$} & \multicolumn{1}{l}{$\textit{ }$} & \multicolumn{4}{l}{$\textsc{O\_{}6408f5}$} & \multicolumn{1}{l}{$\textit{ }$} & \multicolumn{2}{l}{$\textit{\}}\textit{}$}\\
\multicolumn{3}{l}{$\textbf{data}\textit{ }\textsc{O\_{}6408f5}\textit{ }\textit{=}\textit{ }\textsc{O\_{}6408f5}$} & \multicolumn{1}{l}{$\textit{ }$} & \multicolumn{1}{l}{$\textit{\{}$} & \multicolumn{1}{l}{$\textit{ }$} & \multicolumn{1}{l}{$\emph{size}\textit{,}\textit{ }$} & \multicolumn{4}{l}{$\emph{height}\textit{ }$} & \multicolumn{1}{l}{$\textit{::}$} & \multicolumn{1}{l}{$\textit{ }$} & \multicolumn{2}{l}{$\textsc{Int}$} & \multicolumn{2}{l}{$\textit{
}$}\\
\multicolumn{4}{p{25ex}}{                         } & \multicolumn{1}{l}{$\textit{,}$} & \multicolumn{1}{l}{$\textit{ }$} & \multicolumn{5}{l}{$\emph{difficulty}\textit{   }$} & \multicolumn{1}{l}{$\textit{::}$} & \multicolumn{1}{l}{$\textit{ }$} & \multicolumn{3}{l}{$\textsc{Double}$} & \multicolumn{1}{l}{$\textit{
}$}\\
\multicolumn{4}{p{25ex}}{                         } & \multicolumn{1}{l}{$\textit{,}$} & \multicolumn{1}{l}{$\textit{ }$} & \multicolumn{5}{l}{$\emph{previous}\textit{     }$} & \multicolumn{1}{l}{$\textit{::}$} & \multicolumn{1}{l}{$\textit{ }$} & \multicolumn{3}{l}{$\textsc{String}$} & \multicolumn{1}{l}{$\textit{ }\textit{\}}$}\\

\end{tabular}

However, when this object has multiple
keys with values of the same structure,
the best representation is that of a
mapping shown below. This is also an
example of when user may decide to
explicitly add evidence for one of the
alternative representations in the case
when input samples are insufficient.
(like when input samples only contain a
single element dictionary.)

\hypertarget{sec:nonmonotonic-inference}{}
\setlength{\tabcolsep}{1pt}
\begin{tabular}{lllllllllllllllllllllllll}
\multicolumn{4}{l}{$\textbf{data}$} & \multicolumn{1}{l}{$\textit{ }$} & \multicolumn{1}{l}{$\textsc{ExampleMap}$} & \multicolumn{1}{l}{$\textit{ }$} & \multicolumn{1}{l}{$\textit{=}$} & \multicolumn{1}{l}{$\textit{ }$} & \multicolumn{3}{l}{$\textsc{ExampleMap}$} & \multicolumn{1}{l}{$\textit{ }$} & \multicolumn{1}{l}{$\textit{(}$} & \multicolumn{1}{l}{$\textsc{Map}\textit{ }\textsc{Hex}\textit{ }\textsc{ExampleElt}\textit{)}\textit{}$}\\
\multicolumn{4}{l}{$\textbf{data}$} & \multicolumn{1}{l}{$\textit{ }$} & \multicolumn{1}{l}{$\textsc{ExampleElt}$} & \multicolumn{1}{l}{$\textit{ }$} & \multicolumn{1}{l}{$\textit{=}$} & \multicolumn{1}{l}{$\textit{ }$} & \multicolumn{3}{l}{$\textsc{ExampleElt}$} & \multicolumn{1}{l}{$\textit{ }$} & \multicolumn{1}{l}{$\textit{\{}$} & \multicolumn{1}{l}{$\textit{
}$}\\
\multicolumn{4}{p{4ex}}{    } & \multicolumn{2}{l}{$\emph{size}\textit{       }$} & \multicolumn{2}{l}{$\textit{::}$} & \multicolumn{1}{l}{$\textit{ }$} & \multicolumn{1}{l}{$\textsc{Int}$} & \multicolumn{5}{l}{$\textit{
}$}\\
\multicolumn{2}{p{2ex}}{  } & \multicolumn{1}{l}{$\textit{,}$} & \multicolumn{1}{l}{$\textit{ }$} & \multicolumn{2}{l}{$\emph{height}\textit{     }$} & \multicolumn{2}{l}{$\textit{::}$} & \multicolumn{1}{l}{$\textit{ }$} & \multicolumn{1}{l}{$\textsc{Int}$} & \multicolumn{5}{l}{$\textit{
}$}\\
\multicolumn{2}{p{2ex}}{  } & \multicolumn{1}{l}{$\textit{,}$} & \multicolumn{1}{l}{$\textit{ }$} & \multicolumn{2}{l}{$\emph{difficulty}\textit{ }$} & \multicolumn{2}{l}{$\textit{::}$} & \multicolumn{1}{l}{$\textit{ }$} & \multicolumn{2}{l}{$\textsc{Double}$} & \multicolumn{4}{l}{$\textit{
}$}\\
\multicolumn{2}{p{2ex}}{  } & \multicolumn{1}{l}{$\textit{,}$} & \multicolumn{1}{l}{$\textit{ }$} & \multicolumn{2}{l}{$\emph{previous}\textit{   }$} & \multicolumn{2}{l}{$\textit{::}$} & \multicolumn{1}{l}{$\textit{ }$} & \multicolumn{2}{l}{$\textsc{String}$} & \multicolumn{4}{l}{$\textit{ }\textit{\}}$}\\

\end{tabular}

\hypertarget{goal-of-inference}{%
\subsection{Goal of
inference}\label{goal-of-inference}}

Given an undocumented (or incorrectly
labelled) JSON API, we may need to read
the input of Haskell encoding and avoid
checking for the presence of
\emph{unexpected} format deviations. At
the same time, we may decide to accept
all known valid inputs outright so that
we can use types\footnote{Compiler
  feature of checking for unmatched
  cases.} to ensure that the input is
processed exhaustively.

Accordingly, we can assume that the
smallest non-singleton set is a better
approximation type than a singleton set.
We call it \emph{minimal containing set
principle}.

Second, we can prefer types that allow
for a fewer number of \emph{degrees of
freedom} compared with the others, while
conforming to a commonly occurring
structure. We denote it as an
\emph{information content principle}.

Given these principles, and examples of
frequently occurring patterns, we can
infer a reasonable \emph{world of types}
that approximate sets of possible
values. In this way, we can implement
\emph{type system engineering} that
allows deriving type system design
directly from the information about data
structures and the likelihood of their
occurrence.

\hypertarget{problem-definition}{%
\section{Problem
definition}\label{problem-definition}}

As we focus on JSON, we utilize Haskell
encoding of the JSON term for convenient
reading(from Aeson package
{[}\protect\hyperlink{ref-aeson}{1}{]});
specified as follows:

\setlength{\tabcolsep}{1pt}
\begin{tabular}{llllllllllllllllll}
\multicolumn{2}{l}{$\textbf{data}\textit{ }\textsc{Value}\textit{ }$} & \multicolumn{1}{l}{$\textit{=}$} & \multicolumn{1}{l}{$\textit{ }$} & \multicolumn{1}{l}{$\textsc{Object}$} & \multicolumn{1}{l}{$\textit{ }$} & \multicolumn{1}{l}{$\textit{(}\textsc{Map}\textit{ }$} & \multicolumn{1}{l}{$\textsc{String}\textit{ }\textsc{Value}\textit{)}\textit{
}$}\\
\multicolumn{2}{p{11ex}}{           } & \multicolumn{1}{l}{$\textit{|}$} & \multicolumn{1}{l}{$\textit{ }$} & \multicolumn{1}{l}{$\textsc{Array}\textit{ }$} & \multicolumn{1}{l}{$\textit{[}$} & \multicolumn{1}{l}{$\textsc{Value}$} & \multicolumn{1}{l}{$\textit{]}\textit{
}$}\\
\multicolumn{2}{p{11ex}}{           } & \multicolumn{1}{l}{$\textit{|}$} & \multicolumn{1}{l}{$\textit{ }$} & \multicolumn{4}{l}{$\textsc{Null}\textit{                     
}$}\\
\multicolumn{2}{p{11ex}}{           } & \multicolumn{1}{l}{$\textit{|}$} & \multicolumn{1}{l}{$\textit{ }$} & \multicolumn{1}{l}{$\textsc{Number}$} & \multicolumn{1}{l}{$\textit{ }$} & \multicolumn{2}{l}{$\textsc{Scientific}\textit{
}$}\\
\multicolumn{2}{p{11ex}}{           } & \multicolumn{1}{l}{$\textit{|}$} & \multicolumn{1}{l}{$\textit{ }$} & \multicolumn{1}{l}{$\textsc{String}$} & \multicolumn{1}{l}{$\textit{ }$} & \multicolumn{2}{l}{$\textsc{Text}\textit{              
}$}\\
\multicolumn{2}{p{11ex}}{           } & \multicolumn{1}{l}{$\textit{|}$} & \multicolumn{1}{l}{$\textit{ }$} & \multicolumn{4}{l}{$\textsc{Bool}\textit{ }\textsc{Bool}$}\\

\end{tabular}

\hypertarget{defining-type-inference}{%
\subsection{Defining type
inference}\label{defining-type-inference}}

\hypertarget{information-in-the-type-descriptions}{%
\subsubsection{Information in the type
descriptions}\label{information-in-the-type-descriptions}}

If an inference fails, it is always
possible to correct it by introducing an
additional observation (example). To
denote unification operation, or
\textbf{information fusion} between two
type descriptions, we use a
\texttt{Semigroup} interface operation
\texttt{\textless{}\textgreater{}} to
merge types inferred from different
observations. If the semigroup is a
semilattice, then
\texttt{\textless{}\textgreater{}} is
meet operation (least upper bound). Note
that this approach is dual to
traditional unification that
\emph{narrows down} solutions and thus
is join operation (greatest lower
bound). We use a neutral element of the
\texttt{Monoid} to indicate a type
corresponding to no observations.

\setlength{\tabcolsep}{1pt}
\begin{tabular}{llllllllllllllllllllll}
\multicolumn{4}{l}{$\textbf{class}$} & \multicolumn{1}{l}{$\textit{ }$} & \multicolumn{2}{l}{$\textsc{Semigroup}$} & \multicolumn{1}{l}{$\textit{ }$} & \multicolumn{1}{l}{$\emph{ty}$} & \multicolumn{1}{l}{$\textit{ }$} & \multicolumn{1}{l}{$\textbf{where}$} & \multicolumn{1}{l}{$\textit{
}$}\\
\multicolumn{2}{p{2ex}}{  } & \multicolumn{2}{l}{$\textit{(}\diamond$} & \multicolumn{1}{l}{$\textit{)}$} & \multicolumn{2}{l}{$\textit{ }\textit{::}\textit{ }\emph{ty}\textit{ }\textit{->}$} & \multicolumn{1}{l}{$\textit{ }$} & \multicolumn{1}{l}{$\emph{ty}$} & \multicolumn{1}{l}{$\textit{ }$} & \multicolumn{1}{l}{$\textit{->}\textit{ }\emph{ty}$} & \multicolumn{1}{l}{$\textit{}$}\\
\multicolumn{4}{l}{$\textbf{class}$} & \multicolumn{1}{l}{$\textit{ }$} & \multicolumn{2}{l}{$\textsc{Semigroup}$} & \multicolumn{1}{l}{$\textit{ }$} & \multicolumn{1}{l}{$\emph{ty}$} & \multicolumn{3}{l}{$\textit{
}$}\\
\multicolumn{3}{p{3ex}}{   } & \multicolumn{1}{l}{$\textit{=>}$} & \multicolumn{1}{l}{$\textit{ }$} & \multicolumn{1}{l}{$\textsc{Monoid}$} & \multicolumn{2}{l}{$\textit{    }$} & \multicolumn{1}{l}{$\emph{ty}$} & \multicolumn{3}{l}{$\textit{ }\textbf{where}\textit{
}$}\\
\multicolumn{2}{p{2ex}}{  } & \multicolumn{4}{l}{$\emph{mempty}\textit{ }\textit{::}\textit{ }$} & \multicolumn{6}{l}{$\emph{ty}$}\\

\end{tabular}

In other words, we can say that
\texttt{mempty} (or \texttt{bottom})
element corresponds to situation where
\textbf{no information was accepted}
about a possible value (no term was
seen, not even a null). It is a neutral
element of \texttt{Typelike}. For
example, an empty array \texttt{{[}{]}}
can be referred to as an array type with
\texttt{mempty} as an element type. This
represents the view that
\texttt{\textless{}\textgreater{}}
always \textbf{gathers more information}
about the type, as opposed to the
traditional unification that always
\textbf{narrows down} possible
solutions. We describe the laws
described below as QuickCheck
{[}\protect\hyperlink{ref-quickcheck}{4}{]}
properties so that unit testing can be
implemented to detect apparent
violations.

\hypertarget{beyond-set}{%
\subsubsection{Beyond
set}\label{beyond-set}}

In the domain of permissive union types,
a \texttt{beyond} set represents the
case of \textbf{everything permitted} or
a fully dynamic value when we gather the
information that permits every possible
value inside a type. At the first
reading, it may be deemed that a
\texttt{beyond} set should comprise of
only one single element -- the
\texttt{top} one (arriving at complete
bounded semilattice), but this is too
narrow for our purpose of
\emph{monotonically gathering
information}

However, since we defined
\textbf{generalization} operator
\texttt{\textless{}\textgreater{}} as
\textbf{information fusion}
(corresponding to unification in
categorically dual case of strict type
systems.), we may encounter difficulties
in assuring that no information has been
lost during the
generalization\footnote{Examples will be
  provided later.}. Moreover, strict
type systems usually specify more than
one error value, as it should contain
information about error messages and
keep track from where an error has been
originated\footnote{In this case:
  \texttt{beyond\ (Error\ \_)\ =\ True\ \textbar{}\ otherwise\ =\ False}.}.

This observation lets us go well beyond
typing statement of gradual type
inference as a discovery problem from
incomplete information
{[}\protect\hyperlink{ref-gradual-typing}{22}{]}.
Here we consider type inference as a
\textbf{learning problem} furthermore,
find common ground between the dynamic
and the static typing discipline. The
languages relying on the static type
discipline usually consider
\texttt{beyond} as a set of error
messages, as a value should correspond
to a statically assigned and a
\textbf{narrow} type. In this setting,
\texttt{mempty} would be fully
polymorphic type \(\forall{}a. a\).

Languages with dynamic type discipline
will treat \texttt{beyond} as untyped,
dynamic value and \texttt{mempty} again
is an entirely unknown, polymorphic
value (like a type of an element of an
empty array)\footnote{May sound similar
  until we consider adding more
  information to the type.}.

\setlength{\tabcolsep}{1pt}
\begin{tabular}{lllllllllllllllllllll}
\multicolumn{4}{l}{$\textbf{class}$} & \multicolumn{1}{l}{$\textit{ }$} & \multicolumn{1}{l}{$\textit{(}\textsc{Monoid}\textit{ }$} & \multicolumn{1}{l}{$\emph{t}$} & \multicolumn{1}{l}{$\textit{,}$} & \multicolumn{1}{l}{$\textit{ }$} & \multicolumn{1}{l}{$\textsc{Eq}\textit{ }\emph{t}\textit{,}$} & \multicolumn{1}{l}{$\textit{ }\textsc{Show}\textit{ }\emph{t}\textit{)}\textit{
}$}\\
\multicolumn{3}{p{3ex}}{   } & \multicolumn{1}{l}{$\textit{=>}$} & \multicolumn{1}{l}{$\textit{ }$} & \multicolumn{1}{l}{$\textsc{Typelike}$} & \multicolumn{1}{l}{$\textit{ }$} & \multicolumn{1}{l}{$\emph{t}$} & \multicolumn{1}{l}{$\textit{ }$} & \multicolumn{1}{l}{$\textbf{where}$} & \multicolumn{1}{l}{$\textit{
}$}\\
\multicolumn{2}{p{2ex}}{  } & \multicolumn{4}{l}{$\emph{beyond}\textit{ }\textit{::}\textit{ }\emph{t}\textit{ }$} & \multicolumn{2}{l}{$\textit{->}$} & \multicolumn{1}{l}{$\textit{ }$} & \multicolumn{2}{l}{$\textsc{Bool}$}\\

\end{tabular}

Besides, the standard laws for a
\textbf{commutative} \texttt{Monoid}, we
state the new law for the
\texttt{beyond} set: The \texttt{beyond}
set is always \textbf{closed to
information addition} by
\texttt{(\textless{}\textgreater{}a)} or
\texttt{(a\textless{}\textgreater{})}
for any value of \texttt{a}, or
\textbf{submonoid}. In other words, the
\texttt{beyond} set is an attractor of
\texttt{\textless{}\textgreater{}} on
both sides.\footnote{So both for
  ∀\texttt{a(\textless{}\textgreater{}\ a)}
  and
  ∀\texttt{a.(a\textless{}\textgreater{})},
  the result is kept in the
  \texttt{beyond} set.} However, we do
not require \emph{idempotence} of
\texttt{\textless{}\textgreater{}},
which is uniformly present in union type
frameworks based on the lattice
{[}\protect\hyperlink{ref-subtyping-lattice}{25}{]}
and set-based
approaches\footnote{Which use Heyting
  algebras, which have more assumptions
  that the lattice approaches.}{[}\protect\hyperlink{ref-semantic-subtyping}{9}{]}.
Concerning union types, the key property
of the \texttt{beyond} set, is that it
is closed to information acquisition:

In this way, we can specify other
elements of \texttt{beyond} set instead
of a single \texttt{top}. When under
strict type discipline, like that of
Haskell
{[}\protect\hyperlink{ref-GHCZurihac}{21}{]},
we seek to enable each element of the
\texttt{beyond} set to contain at least
one error message\footnote{Note that
  many but not all type constraints are
  semilattice. Please refer to the
  counting example below.}.

We abolish the semilattice requirement
that has been conventionally assumed for
type constraints
{[}\protect\hyperlink{ref-subtype-inequalities}{24}{]},
as this requirement is valid only for
the strict type constraint inference,
not for a more general type inference
considered as a learning problem. As we
observe in example 5 in
sec.~\ref{sec:examples}, we need to
perform a non-monotonic step of choosing
alternative representation after
monotonic steps of merging all the
information.

When a specific instance of
\texttt{Typelike} is not a semilattice
(an idempotent semigroup), we will
explicitly indicate that is the case. It
is convenient validation when testing a
recursive structure of the type. Note
that we abolish semilattice requirement
that was traditionally assumed for type
constraints here
{[}\protect\hyperlink{ref-subtyping-lattice}{25}{]}.
That is because this requirement is
valid only for strict type constraint
inference, not for a more general type
inference as a learning problem. As we
saw on \texttt{ExampleMap} in
sec.~\ref{sec:examples}, we need
non-monotonic inference when dealing
with alternative representations. We
note that this approach significantly
generalized the assumptions compared
with a full lattice subtyping
{[}\protect\hyperlink{ref-subtype-inequalities}{24},
\protect\hyperlink{ref-subtyping-lattice}{25}{]}.

Time to present the \textbf{relation of
typing} and its laws. In order to
preserve proper English word order, we
state that
\(\mathit{ty}\,{}^\backprime{}\mathit{Types}^\backprime{}\,\mathit{val}\)
instead of classical
\(\mathit{val}\mathrm{:}\mathit{ty}\).
Specifying the laws of typing is
important, since we may need to
separately consider the validity of a
domain of types/type constraints, and
that of the sound typing of the terms by
these valid types. The minimal
definition of typing inference relation
and type checking relation is formulated
as consistency between these two
operations.

\setlength{\tabcolsep}{1pt}
\begin{tabular}{lllllllllllllllllllll}
\multicolumn{6}{l}{$\textbf{class}\textit{ }\textsc{Typelike}\textit{ }\emph{ty}\textit{ }$} & \multicolumn{1}{l}{$\textit{=>}\textit{ }$} & \multicolumn{2}{l}{$\emph{ty}\textit{ }$} & \multicolumn{1}{l}{$\textit{`}$} & \multicolumn{1}{l}{$\textsc{Types}\textit{`}\textit{ }\emph{val}\textit{ }\textbf{where}\textit{
}$}\\
\multicolumn{2}{p{3ex}}{   } & \multicolumn{1}{l}{$\emph{infer}$} & \multicolumn{1}{l}{$\textit{ }$} & \multicolumn{1}{l}{$\textit{::}$} & \multicolumn{1}{l}{$\textit{       }$} & \multicolumn{1}{l}{$\emph{val}$} & \multicolumn{1}{l}{$\textit{ }$} & \multicolumn{1}{l}{$\textit{->}$} & \multicolumn{1}{l}{$\textit{ }$} & \multicolumn{1}{l}{$\emph{ty}\textit{
}$}\\
\multicolumn{2}{p{3ex}}{   } & \multicolumn{1}{l}{$\emph{check}$} & \multicolumn{1}{l}{$\textit{ }$} & \multicolumn{1}{l}{$\textit{::}$} & \multicolumn{1}{l}{$\textit{ }\emph{ty}\textit{ }\textit{->}\textit{ }$} & \multicolumn{1}{l}{$\emph{val}$} & \multicolumn{1}{l}{$\textit{ }$} & \multicolumn{1}{l}{$\textit{->}$} & \multicolumn{1}{l}{$\textit{ }$} & \multicolumn{1}{l}{$\textsc{Bool}$}\\

\end{tabular}

First, we note that to describe \emph{no
information}, \texttt{mempty} cannot
correctly type any term. A second
important rule of typing is that all
terms are typed successfully by any
value in the \texttt{beyond} set.
Finally, we state the most intuitive
rule for typing: a type inferred from a
term, must always be valid for that
particular term. The law asserts that
the diagram on the figure commutes:

\begin{figure*}[t]
\begin{center}
% https://tikzcd.yichuanshen.de/#N4Igdg9gJgpgziAXAbVABwnAlgFyxMJZAJgBpiBdUkANwEMAbAVxiRAB12BbOnAC1zAAKgE80MAL4B9AIwAeAHyce-QaPHTiICaXSZc+QigAM5KrUYs2y3gJzCxk2dt0gM2PASIAWM9XrMrIgc3LZqjpoueh6GRGTe5gFWwTaq9kIATiwSUW76nkYkpMaJlkEhKnYOGrIABJx4XPD1oWnVTlo60QZeKL4J-mXWrVUAamWRXXkxvcimAxaBw5WC44HSMtrmMFAA5vBEoABmGRBcSDLUOBBIAMzUAEYwYFB3piBwAkc4SO8MWGBypBASBqHwYHRXsFgawrnQsAw2DDQSAGHQngwAAr5WLBAHYWC5E5nX5XG6Ie6LZIVML2ADG4LpAGtOLUkh0UU8XkhvABOKbE86Id7XO6PZ5QgC0fMGSxSI0EDJgzNZ7I2KLRGOxMyMIAyWF2fB+AtOQrIIFFFJNJMQlwt5Nu1qFdst73ZbEUqSq6g5TqQ5tdsupXvCNXkCg16JgWJxvT1BqNRNNPLJpKp5UUSZtvnt-qD5U4aCwUi01E10e1PV1DBg3yzQpzlrt7vlRecZajMZ1bH1huNrkFSAAbKnbfnlrTgACjjAMjk-YgAKyj80tmltaeznIdrWx3W9xML5e5inj+UremMlnsWoAd1wfAkrJ9Jfrw9HlLXIcvyuvd4fT43i+mw7hWe49gmxoUBIQA
\begin{tikzcd}
                                                                                              &  & \mathit{Type}_1 \times \mathit{Type}_2 \arrow[dd, "<>"] \arrow[rrdd, "\pi_2"] \arrow[lldd, "\pi_1"']                                       &  &                                                                                           \\
                                                                                              &  &                                                                                                                                            &  &                                                                                           \\
\mathit{Type}_1 \arrow[rrdd, phantom, bend left] \arrow[rrdd] \arrow[rr, "<>\mathit{Type}_2"] &  & \mathit{Type}_1<>\mathit{Type}_2 \arrow[dd, "\mathit{check\ value}_2", bend left=49] \arrow[dd, "\mathit{check\ value}_1"', bend right=49] &  & \mathit{Type}_2 \arrow[lldd] \arrow[ll, "\mathit{Type}_1<>"']                             \\
                                                                                              &  &                                                                                                                                            &  &                                                                                           \\
\mathit{Value}_1 \arrow[uu, "\mathit{infer}"] \arrow[rr, "\mathit{check\ with}\ Type_1"']     &  & \mathit{True}                                                                                                                              &  & \mathit{Value}_2 \arrow[uu, "\mathit{infer}"'] \arrow[ll, "\mathit{check\ with}\ Type_2"]
\end{tikzcd}
\end{center}
\caption{Categorical diagram for \texttt{Typelike}.}
\end{figure*}
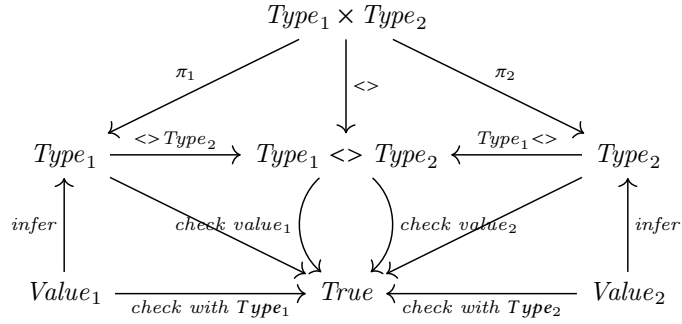

The last law states that the terms are
correctly type-checked after adding more
information into a single type. (For
inference relation, it would be
described as \emph{principal type
property}.) The minimal
\texttt{Typelike} instance is the one
that contains only \texttt{mempty}
corresponding to the case of \emph{no
sample data received}, and a single
\texttt{beyond} element for \emph{all
values permitted}. We will define it
below as \texttt{PresenceConstraint} in
sec.~\ref{sec:presence-absence-constraints}.
These laws are also compatible with the
strict, static type discipline: namely,
the \texttt{beyond} set corresponds to a
set of constraints with at least one
type error, and a task of a compiler to
prevent any program with the terms that
type only to the \texttt{beyond} as a
least upper bound.

\hypertarget{type-engineering-principles}{%
\subsection{Type engineering
principles}\label{type-engineering-principles}}

Considering that we aim to infer a type
from a finite number of samples, we
encounter a \emph{learning problem}, so
we need to use \emph{prior} knowledge
about the domain for inferring types.
Observing that \(a: \text{false}\) we
can expect that in particular cases, we
may obtain that \(a : \text{true}\).
After noting that \(b: 123\), we expect
that \(b: 100\) would also be
acceptable. It means that we need to
consider a typing system to \emph{learn
a reasonable general class from few
instances}. This observation motivates
formulating the type system as an
inference problem. As the purpose is to
deliver the most descriptive\footnote{The
  shortest one according to the
  information complexity principle.}
types, we assume that we need to obtain
a broader view rather than focusing on a
\emph{free type} and applying it to
larger sets whenever it is deemed
justified.

The other principle corresponds to
\textbf{correct operation}. It implies
that having operations regarded on
types, we can find a minimal set of
types that assure correct operation in
the case of unexpected errors. Indeed we
want to apply this theory to infer a
type definition from a finite set of
examples. We also seek to generalize it
to infinite types. We endeavour rules to
be as short as possible. The inference
must also be a \textbf{contravariant}
functor with regards to constructors.
For example, if \texttt{AType\ x\ y}
types \texttt{\{"a":\ X,\ "b":\ Y\}},
then \texttt{x} must type \texttt{X},
and \texttt{y} must type \texttt{Y}.

\hypertarget{constraint-definition}{%
\subsection{Constraint
definition}\label{constraint-definition}}

\hypertarget{flat-type-constraints}{%
\subsubsection{Flat type
constraints}\label{flat-type-constraints}}

Let us first consider typing of flat
type: \texttt{String} (similar treatment
should be given to the
\texttt{Number}.type.)

\setlength{\tabcolsep}{1pt}
\begin{tabular}{lllllllllllllllllll}
\multicolumn{4}{l}{$\textbf{data}$} & \multicolumn{3}{l}{$\textit{ }\textsc{StringConstraint}$} & \multicolumn{1}{l}{$\textit{ }$} & \multicolumn{1}{l}{$\textit{=}\textit{ }\textsc{SCDate}\textit{               }\textit{|}\textit{ }\textsc{SCEmail}\textit{
}$}\\
\multicolumn{2}{p{2ex}}{  } & \multicolumn{1}{l}{$\textit{|}$} & \multicolumn{1}{l}{$\textit{ }$} & \multicolumn{1}{l}{$\textsc{SCEnum}\textit{ }$} & \multicolumn{1}{l}{$\textit{(}$} & \multicolumn{1}{l}{$\textsc{Set}\textit{ }\textsc{Text}\textit{)}$} & \multicolumn{1}{l}{$\textit{ }$} & \multicolumn{1}{l}{$\textit{\{- non-empty set of observed values -\}}\textit{
}$}\\
\multicolumn{2}{p{2ex}}{  } & \multicolumn{1}{l}{$\textit{|}$} & \multicolumn{1}{l}{$\textit{ }$} & \multicolumn{1}{l}{$\textsc{SCNever}$} & \multicolumn{1}{l}{$\textit{ }$} & \multicolumn{3}{l}{$\textit{\{- mempty -\}}\textit{ }\textit{|}\textit{ }\textsc{SCAny}\textit{ }\textit{\{- beyond -\}}$}\\

\end{tabular}

\setlength{\tabcolsep}{1pt}
\begin{tabular}{lllllllllllllllllllllllllllllll}
\multicolumn{4}{l}{$\textbf{instance}$} & \multicolumn{1}{l}{$\textit{ }$} & \multicolumn{7}{l}{$\textsc{StringConstraint}$} & \multicolumn{4}{l}{$\textit{ }\textit{`}\textsc{Types}$} & \multicolumn{1}{l}{$\textit{`}$} & \multicolumn{3}{l}{$\textit{ }\textsc{Text}\textit{ }$} & \multicolumn{1}{l}{$\textbf{where}\textit{
}$}\\
\multicolumn{2}{p{2ex}}{  } & \multicolumn{1}{l}{$\emph{infer}$} & \multicolumn{1}{l}{$\textit{ }$} & \multicolumn{1}{l}{$\textit{(}$} & \multicolumn{4}{l}{$\emph{isValidDate}\textit{  }$} & \multicolumn{2}{l}{$\textit{->}$} & \multicolumn{1}{l}{$\textit{ }$} & \multicolumn{1}{l}{$\textsc{True}$} & \multicolumn{1}{l}{$\textit{)}$} & \multicolumn{1}{l}{$\textit{ }$} & \multicolumn{1}{l}{$\textit{=}$} & \multicolumn{1}{l}{$\textit{ }$} & \multicolumn{3}{l}{$\textsc{SCDate}$} & \multicolumn{1}{l}{$\textit{
}$}\\
\multicolumn{2}{p{2ex}}{  } & \multicolumn{1}{l}{$\emph{infer}$} & \multicolumn{1}{l}{$\textit{ }$} & \multicolumn{1}{l}{$\textit{(}$} & \multicolumn{4}{l}{$\emph{isValidEmail}\textit{ }$} & \multicolumn{2}{l}{$\textit{->}$} & \multicolumn{1}{l}{$\textit{ }$} & \multicolumn{1}{l}{$\textsc{True}$} & \multicolumn{1}{l}{$\textit{)}$} & \multicolumn{1}{l}{$\textit{ }$} & \multicolumn{1}{l}{$\textit{=}$} & \multicolumn{1}{l}{$\textit{ }$} & \multicolumn{4}{l}{$\textsc{SCEmail}\textit{
}$}\\
\multicolumn{2}{p{2ex}}{  } & \multicolumn{1}{l}{$\emph{infer}$} & \multicolumn{2}{l}{$\textit{  }$} & \multicolumn{10}{l}{$\emph{value}\textit{                 }$} & \multicolumn{1}{l}{$\textit{=}$} & \multicolumn{1}{l}{$\textit{ }$} & \multicolumn{4}{l}{$\textsc{SCEnum}\textit{
}$}\\
\multicolumn{15}{p{31ex}}{                               } & \multicolumn{1}{l}{$\textit{\$}$} & \multicolumn{1}{l}{$\textit{ }$} & \multicolumn{4}{l}{$\emph{Set.singleton}\textit{ }\emph{value}\textit{
}$}\\
\multicolumn{2}{p{2ex}}{  } & \multicolumn{13}{l}{$\emph{infer}\textit{  }\textit{\_{}}\textit{                     }$} & \multicolumn{1}{l}{$\textit{=}$} & \multicolumn{1}{l}{$\textit{ }$} & \multicolumn{4}{l}{$\textsc{SCAny}\textit{
}$}\\
\multicolumn{21}{l}{$\textit{
}$}\\
\multicolumn{2}{p{2ex}}{  } & \multicolumn{1}{l}{$\emph{check}$} & \multicolumn{2}{l}{$\textit{  }$} & \multicolumn{2}{l}{$\textsc{SCDate}\textit{     }$} & \multicolumn{1}{l}{$\emph{s}$} & \multicolumn{1}{l}{$\textit{ }$} & \multicolumn{1}{l}{$\textit{=}$} & \multicolumn{1}{l}{$\textit{ }$} & \multicolumn{7}{l}{$\emph{isValidDate}\textit{  }$} & \multicolumn{1}{l}{$\emph{s}$} & \multicolumn{2}{l}{$\textit{
}$}\\
\multicolumn{2}{p{2ex}}{  } & \multicolumn{1}{l}{$\emph{check}$} & \multicolumn{2}{l}{$\textit{  }$} & \multicolumn{1}{l}{$\textsc{SCEmail}$} & \multicolumn{1}{l}{$\textit{    }$} & \multicolumn{1}{l}{$\emph{s}$} & \multicolumn{1}{l}{$\textit{ }$} & \multicolumn{1}{l}{$\textit{=}$} & \multicolumn{1}{l}{$\textit{ }$} & \multicolumn{7}{l}{$\emph{isValidEmail}\textit{ }$} & \multicolumn{1}{l}{$\emph{s}$} & \multicolumn{2}{l}{$\textit{
}$}\\
\multicolumn{2}{p{2ex}}{  } & \multicolumn{1}{l}{$\emph{check}$} & \multicolumn{2}{l}{$\textit{ }\textit{(}$} & \multicolumn{1}{l}{$\textsc{SCEnum}\textit{ }$} & \multicolumn{1}{l}{$\emph{vs}\textit{)}\textit{ }$} & \multicolumn{1}{l}{$\emph{s}$} & \multicolumn{1}{l}{$\textit{ }$} & \multicolumn{1}{l}{$\textit{=}$} & \multicolumn{1}{l}{$\textit{ }$} & \multicolumn{7}{l}{$\emph{s}\textit{ }\textit{`}\emph{Set.member}$} & \multicolumn{1}{l}{$\textit{`}$} & \multicolumn{2}{l}{$\textit{ }\emph{vs}\textit{
}$}\\
\multicolumn{2}{p{2ex}}{  } & \multicolumn{1}{l}{$\emph{check}$} & \multicolumn{2}{l}{$\textit{  }$} & \multicolumn{1}{l}{$\textsc{SCNever}$} & \multicolumn{1}{l}{$\textit{    }$} & \multicolumn{1}{l}{$\textit{\_{}}$} & \multicolumn{1}{l}{$\textit{ }$} & \multicolumn{1}{l}{$\textit{=}$} & \multicolumn{1}{l}{$\textit{ }$} & \multicolumn{10}{l}{$\textsc{False}\textit{
}$}\\
\multicolumn{2}{p{2ex}}{  } & \multicolumn{1}{l}{$\emph{check}$} & \multicolumn{2}{l}{$\textit{  }$} & \multicolumn{2}{l}{$\textsc{SCAny}\textit{      }$} & \multicolumn{1}{l}{$\textit{\_{}}$} & \multicolumn{1}{l}{$\textit{ }$} & \multicolumn{1}{l}{$\textit{=}$} & \multicolumn{1}{l}{$\textit{ }$} & \multicolumn{10}{l}{$\textsc{True}$}\\

\end{tabular}

\setlength{\tabcolsep}{1pt}
\begin{tabular}{llllllllllllllllllllllllll}
\multicolumn{4}{l}{$\textbf{instance}\textit{ }$} & \multicolumn{6}{l}{$\textsc{Semigroup}$} & \multicolumn{6}{l}{$\textit{ }\textsc{StringConstraint}\textit{ }\textbf{where}\textit{
}$}\\
\multicolumn{2}{p{2ex}}{  } & \multicolumn{2}{l}{$\textsc{SCNever}$} & \multicolumn{2}{l}{$\textit{    }$} & \multicolumn{1}{l}{$\diamond$} & \multicolumn{2}{l}{$\textit{  }$} & \multicolumn{1}{l}{$\alpha$} & \multicolumn{3}{l}{$\textit{                   }$} & \multicolumn{1}{l}{$\textit{=}$} & \multicolumn{1}{l}{$\textit{ }$} & \multicolumn{1}{l}{$\alpha\textit{
}$}\\
\multicolumn{2}{p{2ex}}{  } & \multicolumn{4}{l}{$\textsc{SCAny}\textit{      }$} & \multicolumn{1}{l}{$\diamond$} & \multicolumn{2}{l}{$\textit{  }$} & \multicolumn{1}{l}{$\textit{\_{}}$} & \multicolumn{3}{l}{$\textit{                   }$} & \multicolumn{1}{l}{$\textit{=}$} & \multicolumn{1}{l}{$\textit{ }$} & \multicolumn{1}{l}{$\textsc{SCAny}\textit{
}$}\\
\multicolumn{2}{p{2ex}}{  } & \multicolumn{4}{l}{$\textsc{SCDate}\textit{     }$} & \multicolumn{1}{l}{$\diamond$} & \multicolumn{2}{l}{$\textit{  }$} & \multicolumn{4}{l}{$\textsc{SCDate}\textit{              }$} & \multicolumn{1}{l}{$\textit{=}$} & \multicolumn{1}{l}{$\textit{ }$} & \multicolumn{1}{l}{$\textsc{SCDate}\textit{
}$}\\
\multicolumn{2}{p{2ex}}{  } & \multicolumn{2}{l}{$\textsc{SCEmail}$} & \multicolumn{2}{l}{$\textit{    }$} & \multicolumn{1}{l}{$\diamond$} & \multicolumn{2}{l}{$\textit{  }$} & \multicolumn{2}{l}{$\textsc{SCEmail}$} & \multicolumn{2}{l}{$\textit{             }$} & \multicolumn{1}{l}{$\textit{=}$} & \multicolumn{1}{l}{$\textit{ }$} & \multicolumn{1}{l}{$\textsc{SCEmail}\textit{
}$}\\
\multicolumn{2}{p{2ex}}{  } & \multicolumn{2}{l}{$\textit{(}\textsc{SCEnum}$} & \multicolumn{1}{l}{$\textit{ }\alpha\textit{)}$} & \multicolumn{1}{l}{$\textit{ }$} & \multicolumn{1}{l}{$\diamond$} & \multicolumn{1}{l}{$\textit{ }$} & \multicolumn{1}{l}{$\textit{(}$} & \multicolumn{2}{l}{$\textsc{SCEnum}\textit{ }$} & \multicolumn{1}{l}{$\beta\textit{)}$} & \multicolumn{4}{l}{$\textit{
}$}\\
\multicolumn{3}{p{4ex}}{    } & \multicolumn{2}{l}{$\textit{|}\textit{ }\emph{length}$} & \multicolumn{1}{l}{$\textit{ }$} & \multicolumn{1}{l}{$\textit{(}\alpha$} & \multicolumn{1}{l}{$\textit{ }$} & \multicolumn{1}{l}{$\textit{`}$} & \multicolumn{3}{l}{$\emph{Set.union}$} & \multicolumn{1}{l}{$\textit{`}\textit{ }\beta\textit{)}\textit{ }\leq\textit{ }10\textit{ }$} & \multicolumn{1}{l}{$\textit{=}$} & \multicolumn{1}{l}{$\textit{ }$} & \multicolumn{1}{l}{$\textsc{SCEnum}\textit{ }\textit{(}\alpha\textit{ }\diamond\textit{ }\beta\textit{)}\textit{
}$}\\
\multicolumn{2}{p{2ex}}{  } & \multicolumn{4}{l}{$\textit{\_{}}\textit{          }$} & \multicolumn{1}{l}{$\diamond$} & \multicolumn{2}{l}{$\textit{  }$} & \multicolumn{4}{l}{$\textit{\_{}}\textit{                   }$} & \multicolumn{1}{l}{$\textit{=}$} & \multicolumn{1}{l}{$\textit{ }$} & \multicolumn{1}{l}{$\textsc{SCAny}$}\\

\end{tabular}

\hypertarget{free-union-type}{%
\subsubsection{Free union
type}\label{free-union-type}}

Before we endeavour on finding type
constraints for compound values (arrays
and objects), it might be instructive to
find a notion of \emph{free type}, that
is a type with no additional laws but
the ones stated above. Given a term with
arbitrary constructors we can infer a
\emph{free type} for every term set
\(T\) as follows: For any \(T\) value
type \(\mathit{Set}\ T\) satisfies our
notion of \emph{free type} specified as
follows:

\setlength{\tabcolsep}{1pt}
\begin{tabular}{llllllllllll}
\multicolumn{1}{l}{$\textbf{data}\textit{ }\textsc{FreeType}\textit{ }\alpha\textit{ }\textit{=}\textit{ }\textsc{FreeType}\textit{ }\textit{\{}\textit{ }\emph{captured}\textit{ }\textit{::}\textit{ }\textsc{Set}\textit{ }\alpha\textit{ }\textit{\}}\textit{ }\textit{|}\textit{ }\textsc{Full}$}\\

\end{tabular}

\setlength{\tabcolsep}{1pt}
\begin{tabular}{lllllllllllllllllllllllllllllll}
\multicolumn{7}{l}{$\textbf{instance}$} & \multicolumn{1}{l}{$\textit{ }$} & \multicolumn{6}{l}{$\textit{(}\textsc{Ord}\textit{ }\alpha\textit{,}\textit{ }\textsc{Eq}$} & \multicolumn{1}{l}{$\textit{ }$} & \multicolumn{7}{l}{$\alpha\textit{)}\textit{
}$}\\
\multicolumn{4}{p{6ex}}{      } & \multicolumn{3}{l}{$\textit{=>}$} & \multicolumn{1}{l}{$\textit{ }$} & \multicolumn{6}{l}{$\textsc{Semigroup}\textit{ }$} & \multicolumn{1}{l}{$\textit{(}$} & \multicolumn{7}{l}{$\textsc{FreeType}\textit{ }\alpha\textit{)}\textit{ }\textbf{where}\textit{
}$}\\
\multicolumn{2}{p{2ex}}{  } & \multicolumn{2}{l}{$\textsc{Full}$} & \multicolumn{1}{l}{$\textit{ }$} & \multicolumn{3}{c}{$\diamond$} & \multicolumn{1}{l}{$\textit{ }$} & \multicolumn{1}{l}{$\textit{\_{}}\textit{    }$} & \multicolumn{1}{l}{$\textit{=}$} & \multicolumn{1}{l}{$\textit{ }$} & \multicolumn{4}{l}{$\textsc{Full}$} & \multicolumn{6}{l}{$\textit{
}$}\\
\multicolumn{2}{p{2ex}}{  } & \multicolumn{1}{l}{$\textit{\_{}}$} & \multicolumn{2}{l}{$\textit{    }$} & \multicolumn{3}{c}{$\diamond$} & \multicolumn{1}{l}{$\textit{ }$} & \multicolumn{1}{l}{$\textsc{Full}\textit{ }$} & \multicolumn{1}{l}{$\textit{=}$} & \multicolumn{1}{l}{$\textit{ }$} & \multicolumn{4}{l}{$\textsc{Full}$} & \multicolumn{6}{l}{$\textit{
}$}\\
\multicolumn{2}{p{2ex}}{  } & \multicolumn{1}{l}{$\alpha$} & \multicolumn{2}{l}{$\textit{    }$} & \multicolumn{3}{c}{$\diamond$} & \multicolumn{1}{l}{$\textit{ }$} & \multicolumn{1}{l}{$\beta\textit{    }$} & \multicolumn{1}{l}{$\textit{=}$} & \multicolumn{1}{l}{$\textit{ }$} & \multicolumn{10}{l}{$\textsc{FreeType}\textit{
}$}\\
\multicolumn{10}{p{15ex}}{               } & \multicolumn{1}{l}{$\textit{\$}$} & \multicolumn{1}{l}{$\textit{ }$} & \multicolumn{5}{l}{$\textit{(}\emph{Set.union}$} & \multicolumn{1}{l}{$\textit{ }$} & \multicolumn{1}{l}{$\textit{`}$} & \multicolumn{3}{l}{$\emph{on}\textit{`}\textit{ }\emph{captured}\textit{)}\textit{ }\alpha\textit{ }\beta\textit{}$}\\
\multicolumn{7}{l}{$\textbf{instance}$} & \multicolumn{1}{l}{$\textit{ }$} & \multicolumn{2}{l}{$\textit{(}\textsc{Ord}\textit{ }\alpha$} & \multicolumn{1}{l}{$\textit{,}$} & \multicolumn{1}{l}{$\textit{ }$} & \multicolumn{2}{l}{$\textsc{Eq}$} & \multicolumn{3}{l}{$\textit{ }\alpha\textit{,}\textit{ }\textsc{Show}$} & \multicolumn{1}{l}{$\textit{ }$} & \multicolumn{1}{l}{$\alpha$} & \multicolumn{1}{l}{$\textit{)}$} & \multicolumn{2}{l}{$\textit{
}$}\\
\multicolumn{4}{p{6ex}}{      } & \multicolumn{3}{l}{$\textit{=>}$} & \multicolumn{1}{l}{$\textit{ }$} & \multicolumn{4}{l}{$\textsc{Typelike}$} & \multicolumn{1}{l}{$\textit{ }$} & \multicolumn{1}{l}{$\textit{(}$} & \multicolumn{3}{l}{$\textsc{FreeType}$} & \multicolumn{1}{l}{$\textit{ }$} & \multicolumn{1}{l}{$\alpha$} & \multicolumn{1}{l}{$\textit{)}$} & \multicolumn{2}{l}{$\textit{ }\textbf{where}\textit{
}$}\\
\multicolumn{2}{p{2ex}}{  } & \multicolumn{5}{l}{$\emph{beyond}$} & \multicolumn{1}{l}{$\textit{ }$} & \multicolumn{5}{l}{$\textit{=}\textit{ }\textit{(}\textit{==}\textsc{Full}$} & \multicolumn{1}{l}{$\textit{)}$} & \multicolumn{8}{l}{$\textit{
}$}\\
\multicolumn{22}{l}{$\textit{}$}\\
\multicolumn{7}{l}{$\textbf{instance}$} & \multicolumn{1}{l}{$\textit{ }$} & \multicolumn{4}{l}{$\textit{(}\textsc{Ord}\textit{ }\alpha\textit{,}\textit{ }$} & \multicolumn{2}{l}{$\textsc{Eq}$} & \multicolumn{1}{l}{$\textit{ }$} & \multicolumn{1}{l}{$\alpha$} & \multicolumn{1}{l}{$\textit{,}\textit{ }\textsc{Show}$} & \multicolumn{1}{l}{$\textit{ }$} & \multicolumn{1}{l}{$\alpha$} & \multicolumn{1}{l}{$\textit{)}$} & \multicolumn{2}{l}{$\textit{
}$}\\
\multicolumn{4}{p{6ex}}{      } & \multicolumn{3}{l}{$\textit{=>}$} & \multicolumn{1}{l}{$\textit{ }$} & \multicolumn{4}{l}{$\textsc{FreeType}$} & \multicolumn{2}{l}{$\textit{ }\alpha$} & \multicolumn{1}{l}{$\textit{ }$} & \multicolumn{1}{l}{$\textit{`}$} & \multicolumn{1}{l}{$\textsc{Types}\textit{`}$} & \multicolumn{1}{l}{$\textit{ }$} & \multicolumn{1}{l}{$\alpha$} & \multicolumn{1}{l}{$\textit{ }$} & \multicolumn{2}{l}{$\textbf{where}\textit{
}$}\\
\multicolumn{22}{l}{$\textit{
}$}\\
\multicolumn{2}{p{2ex}}{  } & \multicolumn{3}{l}{$\emph{infer}$} & \multicolumn{12}{l}{$\textit{                    }$} & \multicolumn{1}{l}{$\textit{=}$} & \multicolumn{1}{l}{$\textit{ }$} & \multicolumn{3}{l}{$\textsc{FreeType}\textit{ }\textit{.}\textit{ }\emph{Set.singleton}\textit{
}$}\\
\multicolumn{2}{p{2ex}}{  } & \multicolumn{3}{l}{$\emph{check}$} & \multicolumn{2}{l}{$\textit{ }$} & \multicolumn{9}{l}{$\textsc{Full}\textit{         }$} & \multicolumn{1}{l}{$\emph{\_{}term}\textit{ }$} & \multicolumn{1}{l}{$\textit{=}$} & \multicolumn{1}{l}{$\textit{ }$} & \multicolumn{2}{l}{$\textsc{True}$} & \multicolumn{1}{l}{$\textit{
}$}\\
\multicolumn{2}{p{2ex}}{  } & \multicolumn{3}{l}{$\emph{check}$} & \multicolumn{2}{l}{$\textit{ }$} & \multicolumn{9}{l}{$\textit{(}\textsc{FreeType}\textit{ }\emph{s}\textit{)}\textit{ }$} & \multicolumn{1}{l}{$\emph{term}\textit{  }$} & \multicolumn{1}{l}{$\textit{=}$} & \multicolumn{1}{l}{$\textit{ }$} & \multicolumn{2}{l}{$\emph{term}$} & \multicolumn{1}{l}{$\textit{ }\textit{`}\emph{Set.member}\textit{`}\textit{ }\emph{s}$}\\

\end{tabular}

This definition is deemed sound and
applicable to finite sets of terms or
values. For a set of values:
\texttt{{[}"yes",\ "no",\ "error"{]}},
we may reasonably consider that type is
an appropriate approximation of C-style
enumeration, or Haskell-style ADT
without constructor arguments. However,
the deficiency of this notion of
\emph{free type} is that it does not
allow generalizing in infinite and
recursive domains! It only allows to
utilize objects from the sample.

\hypertarget{sec:presence-absence-constraints}{%
\subsubsection{Presence and absence
constraint}\label{sec:presence-absence-constraints}}

We call the degenerate case of
\texttt{Typelike} a \emph{presence or
absence constraint}. It just checks that
the type contains at least one
observation of the input value or no
observations at all. It is vital as it
can be used to specify an element type
of an empty array. After seeing
\texttt{true} value, we also expect
\texttt{false}, so we can say that it is
also a primary constraint for
pragmatically indivisible like the set
of boolean values. The same observation
is valid for \texttt{null} values, as
there is only one \texttt{null} value
ever to observe.

\setlength{\tabcolsep}{1pt}
\begin{tabular}{llllllllllllllllllll}
\multicolumn{1}{l}{$\textbf{type}$} & \multicolumn{1}{l}{$\textit{ }$} & \multicolumn{1}{l}{$\textsc{BoolConstraint}$} & \multicolumn{1}{l}{$\textit{ }$} & \multicolumn{1}{l}{$\textit{=}$} & \multicolumn{1}{l}{$\textit{ }$} & \multicolumn{1}{l}{$\textsc{PresenceConstraint}$} & \multicolumn{1}{l}{$\textit{ }$} & \multicolumn{1}{l}{$\textsc{Bool}\textit{}$}\\
\multicolumn{1}{l}{$\textbf{type}$} & \multicolumn{1}{l}{$\textit{ }$} & \multicolumn{1}{l}{$\textsc{NullConstraint}$} & \multicolumn{1}{l}{$\textit{ }$} & \multicolumn{1}{l}{$\textit{=}$} & \multicolumn{1}{l}{$\textit{ }$} & \multicolumn{1}{l}{$\textsc{PresenceConstraint}$} & \multicolumn{1}{l}{$\textit{ }$} & \multicolumn{1}{l}{$\textit{(}\textit{)}\textit{}$}\\
\multicolumn{1}{l}{$\textbf{data}$} & \multicolumn{1}{l}{$\textit{ }$} & \multicolumn{7}{l}{$\textsc{PresenceConstraint}\textit{ }\alpha\textit{ }\textit{=}\textit{ }\textsc{Present}\textit{ }\textit{|}\textit{ }\textsc{Absent}$}\\

\end{tabular}

\hypertarget{variants}{%
\paragraph{Variants}\label{variants}}

It is simple to represent a variant of
two \emph{mutually exclusive} types.
They can be implemented with a type
related to \texttt{Either} type that
assumes these types are exclusive, we
denote it by \texttt{:\textbar{}:}. In
other words for
\texttt{Int\ :\textbar{}:\ String} type,
we first control whether the value is an
\texttt{Int}, and if this check fails,
we attempt to check it as a
\texttt{String}. Variant records are
slightly more complicated, as it may be
unclear which typing is better to use:

\begin{Shaded}
\begin{Highlighting}[]
\FunctionTok{\{}\DataTypeTok{"message"}\FunctionTok{:} \StringTok{"Where can I submit my proposal?"}\FunctionTok{,} \DataTypeTok{"uid"} \FunctionTok{:} \DecValTok{1014}\FunctionTok{\}}
\FunctionTok{\{}\DataTypeTok{"error"}  \FunctionTok{:} \StringTok{"Authorization failed"}\FunctionTok{,}            \DataTypeTok{"code"}\FunctionTok{:}  \DecValTok{401}\FunctionTok{\}}
\end{Highlighting}
\end{Shaded}

\setlength{\tabcolsep}{1pt}
\begin{tabular}{llllllllllllllllllllll}
\multicolumn{4}{l}{$\textbf{data}\textit{ }\textsc{OurRecord}$} & \multicolumn{1}{l}{$\textit{ }$} & \multicolumn{1}{l}{$\textit{=}$} & \multicolumn{6}{l}{$\textit{ }\textsc{OurRecord}\textit{ }\textit{\{}\textit{
}$}\\
\multicolumn{3}{p{7ex}}{       } & \multicolumn{1}{l}{$\emph{message}$} & \multicolumn{1}{l}{$\textit{,}$} & \multicolumn{1}{l}{$\textit{ }$} & \multicolumn{1}{l}{$\emph{error}\textit{ }$} & \multicolumn{1}{l}{$\textit{::}$} & \multicolumn{1}{l}{$\textit{ }$} & \multicolumn{1}{l}{$\textsc{Maybe}$} & \multicolumn{1}{l}{$\textit{ }$} & \multicolumn{1}{l}{$\textsc{String}\textit{
}$}\\
\multicolumn{2}{p{5ex}}{     } & \multicolumn{1}{l}{$\textit{,}\textit{ }$} & \multicolumn{3}{l}{$\emph{code}\textit{,}\textit{    }$} & \multicolumn{1}{l}{$\emph{uid}\textit{   }$} & \multicolumn{1}{l}{$\textit{::}$} & \multicolumn{1}{l}{$\textit{ }$} & \multicolumn{1}{l}{$\textsc{Maybe}$} & \multicolumn{1}{l}{$\textit{ }$} & \multicolumn{1}{l}{$\textsc{Int}\textit{ }\textit{\}}$}\\

\end{tabular}

\setlength{\tabcolsep}{1pt}
\begin{tabular}{llllllllllllllllllllllllllll}
\multicolumn{4}{l}{$\textbf{data}\textit{ }\textsc{OurRecord2}\textit{ }$} & \multicolumn{1}{l}{$\textit{=}$} & \multicolumn{13}{l}{$\textit{
}$}\\
\multicolumn{3}{p{7ex}}{       } & \multicolumn{1}{l}{$\textsc{Message}\textit{  }$} & \multicolumn{1}{l}{$\textit{\{}$} & \multicolumn{1}{l}{$\textit{ }$} & \multicolumn{1}{l}{$\emph{message}\textit{ }$} & \multicolumn{1}{l}{$\textit{::}$} & \multicolumn{1}{l}{$\textit{ }$} & \multicolumn{1}{l}{$\textsc{String}$} & \multicolumn{1}{l}{$\textit{,}$} & \multicolumn{1}{l}{$\textit{ }$} & \multicolumn{1}{l}{$\emph{uid}\textit{  }$} & \multicolumn{1}{l}{$\textit{::}$} & \multicolumn{1}{l}{$\textit{ }$} & \multicolumn{1}{l}{$\textsc{Int}$} & \multicolumn{1}{l}{$\textit{ }$} & \multicolumn{1}{l}{$\textit{\}}\textit{
}$}\\
\multicolumn{2}{p{5ex}}{     } & \multicolumn{1}{l}{$\textit{|}\textit{ }$} & \multicolumn{1}{l}{$\textsc{Error}\textit{    }$} & \multicolumn{1}{l}{$\textit{\{}$} & \multicolumn{1}{l}{$\textit{ }$} & \multicolumn{1}{l}{$\emph{error}\textit{   }$} & \multicolumn{1}{l}{$\textit{::}$} & \multicolumn{1}{l}{$\textit{ }$} & \multicolumn{1}{l}{$\textsc{String}$} & \multicolumn{1}{l}{$\textit{,}$} & \multicolumn{1}{l}{$\textit{ }$} & \multicolumn{1}{l}{$\emph{code}\textit{ }$} & \multicolumn{1}{l}{$\textit{::}$} & \multicolumn{1}{l}{$\textit{ }$} & \multicolumn{1}{l}{$\textsc{Int}$} & \multicolumn{1}{l}{$\textit{ }$} & \multicolumn{1}{l}{$\textit{\}}$}\\

\end{tabular}

The best attempt here is to rely on the
available examples being reasonably
exhaustive. That is, we can estimate how
many examples we have for each, and how
many of them match. Then, we compare
this number with type complexity (with
options being more complex to process
because they need additional
\texttt{case} expression.) In such
cases, the latter definition has only
one \texttt{Maybe} field (on the
toplevel optionality is one), while the
former definition has four
\texttt{Maybe} fields (optionality is
four). When we obtain more samples, the
pattern emerges:

\begin{Shaded}
\begin{Highlighting}[]
\FunctionTok{\{}\DataTypeTok{"error"}  \FunctionTok{:} \StringTok{"Authorization failed"}\FunctionTok{,}            \DataTypeTok{"code"}\FunctionTok{:}  \DecValTok{401}\FunctionTok{\}}
\FunctionTok{\{}\DataTypeTok{"message"}\FunctionTok{:} \StringTok{"Where can I submit my proposal?"}\FunctionTok{,} \DataTypeTok{"uid"} \FunctionTok{:} \DecValTok{1014}\FunctionTok{\}}
\FunctionTok{\{}\DataTypeTok{"message"}\FunctionTok{:} \StringTok{"Sent it to HotCRP"}\FunctionTok{,}               \DataTypeTok{"uid"} \FunctionTok{:}   \DecValTok{93}\FunctionTok{\}}
\FunctionTok{\{}\DataTypeTok{"message"}\FunctionTok{:} \StringTok{"Thanks!"}\FunctionTok{,}                         \DataTypeTok{"uid"} \FunctionTok{:} \DecValTok{1014}\FunctionTok{\}}
\FunctionTok{\{}\DataTypeTok{"error"}  \FunctionTok{:} \StringTok{"Missing user"}\FunctionTok{,}                    \DataTypeTok{"code"}\FunctionTok{:}  \DecValTok{404}\FunctionTok{\}}
\end{Highlighting}
\end{Shaded}

\hypertarget{type-cost-function}{%
\paragraph{Type cost
function}\label{type-cost-function}}

Since we are interested in types with
less complexity and less optionality, we
will define cost function as follows:

\setlength{\tabcolsep}{1pt}
\begin{tabular}{lllllllllllllllllllllllllllllllllll}
\multicolumn{8}{l}{$\textbf{class}\textit{ }\textsc{Typelike}\textit{ }$} & \multicolumn{1}{l}{$\emph{ty}$} & \multicolumn{1}{l}{$\textit{ }$} & \multicolumn{2}{l}{$\textit{=>}$} & \multicolumn{1}{l}{$\textit{ }$} & \multicolumn{12}{l}{$\textsc{TypeCost}\textit{ }\emph{ty}\textit{ }\textbf{where}\textit{
}$}\\
\multicolumn{2}{p{2ex}}{  } & \multicolumn{3}{l}{$\emph{typeCost}$} & \multicolumn{1}{l}{$\textit{ }$} & \multicolumn{1}{l}{$\textit{::}$} & \multicolumn{1}{l}{$\textit{  }$} & \multicolumn{1}{l}{$\emph{ty}$} & \multicolumn{1}{l}{$\textit{ }$} & \multicolumn{2}{l}{$\textit{->}$} & \multicolumn{1}{l}{$\textit{ }$} & \multicolumn{12}{l}{$\textsc{TyCost}\textit{
}$}\\
\multicolumn{2}{p{2ex}}{  } & \multicolumn{3}{l}{$\emph{typeCost}$} & \multicolumn{1}{l}{$\textit{ }$} & \multicolumn{1}{l}{$\alpha\textit{ }$} & \multicolumn{1}{l}{$\textit{=}\textit{ }$} & \multicolumn{1}{l}{$\textbf{if}$} & \multicolumn{1}{l}{$\textit{ }$} & \multicolumn{1}{l}{$\alpha$} & \multicolumn{1}{l}{$\textit{ }$} & \multicolumn{6}{l}{$\textit{==}\textit{ }\emph{mempty}\textit{ }\textbf{then}\textit{ }$} & \multicolumn{1}{l}{$0$} & \multicolumn{3}{l}{$\textit{ }\textbf{else}$} & \multicolumn{1}{l}{$\textit{ }$} & \multicolumn{1}{l}{$1$} & \multicolumn{1}{l}{$\textit{}$}\\
\multicolumn{3}{l}{$\textbf{instance}$} & \multicolumn{1}{l}{$\textit{ }$} & \multicolumn{6}{l}{$\textsc{Semigroup}$} & \multicolumn{1}{l}{$\textit{ }$} & \multicolumn{3}{l}{$\textsc{TyCost}$} & \multicolumn{1}{l}{$\textit{ }$} & \multicolumn{1}{l}{$\textbf{where}$} & \multicolumn{1}{l}{$\textit{ }$} & \multicolumn{1}{l}{$\textit{(}\diamond$} & \multicolumn{1}{l}{$\textit{)}$} & \multicolumn{1}{l}{$\textit{   }$} & \multicolumn{1}{l}{$\textit{=}$} & \multicolumn{1}{l}{$\textit{ }$} & \multicolumn{1}{l}{$\textit{(}$} & \multicolumn{1}{l}{$\textit{+}$} & \multicolumn{1}{l}{$\textit{)}\textit{}$}\\
\multicolumn{3}{l}{$\textbf{instance}$} & \multicolumn{1}{l}{$\textit{ }$} & \multicolumn{7}{l}{$\textsc{Monoid}\textit{    }$} & \multicolumn{3}{l}{$\textsc{TyCost}$} & \multicolumn{1}{l}{$\textit{ }$} & \multicolumn{1}{l}{$\textbf{where}$} & \multicolumn{1}{l}{$\textit{ }$} & \multicolumn{3}{l}{$\emph{mempty}\textit{ }$} & \multicolumn{1}{l}{$\textit{=}$} & \multicolumn{1}{l}{$\textit{ }$} & \multicolumn{1}{l}{$0$} & \multicolumn{2}{l}{$\textit{
}$}\\
\multicolumn{25}{l}{$\textit{}$}\\
\multicolumn{25}{l}{$\textbf{newtype}\textit{ }\textsc{TyCost}\textit{ }\textit{=}\textit{ }\textsc{TyCost}\textit{ }\textsc{Int}$}\\

\end{tabular}

When presented with several alternate
representations from the same set of
observations, we will use this function
to select the least complex
representation of the type. For flat
constraints as above, we infer that they
offer no optionality when no
observations occurred (cost of
\texttt{mempty} is \(0\)), otherwise,
the cost is \(1\). Type cost should be
non-negative, and non-decreasing when we
add new observations to the type.

\hypertarget{object-constraint}{%
\subsubsection{Object
constraint}\label{object-constraint}}

To avoid information loss, a constraint
for JSON object type is introduced in
such a way to \textbf{simultaneously
gather information} about representing
it either as a \texttt{Map}, or a
record. The typing of \texttt{Map} would
be specified as follows, with the
optionality cost being a sum of
optionalities in its fields.

\setlength{\tabcolsep}{1pt}
\begin{tabular}{lllllllllllllllllllllllllllllll}
\multicolumn{4}{l}{$\textbf{data}$} & \multicolumn{1}{l}{$\textit{ }$} & \multicolumn{7}{l}{$\textsc{MappingConstraint}$} & \multicolumn{1}{l}{$\textit{ }$} & \multicolumn{1}{l}{$\textit{=}$} & \multicolumn{7}{l}{$\textit{ }\textsc{MappingNever}\textit{ }\textit{-- mempty}\textit{
}$}\\
\multicolumn{3}{p{3ex}}{   } & \multicolumn{1}{l}{$\textit{|}$} & \multicolumn{1}{l}{$\textit{ }$} & \multicolumn{7}{l}{$\textsc{MappingConstraint}$} & \multicolumn{1}{l}{$\textit{ }$} & \multicolumn{1}{l}{$\textit{\{}$} & \multicolumn{7}{l}{$\textit{
}$}\\
\multicolumn{7}{p{7ex}}{       } & \multicolumn{6}{l}{$\emph{keyConstraint}\textit{   }$} & \multicolumn{2}{l}{$\textit{::}$} & \multicolumn{1}{l}{$\textit{ }$} & \multicolumn{5}{l}{$\textsc{StringConstraint}\textit{
}$}\\
\multicolumn{5}{p{5ex}}{     } & \multicolumn{2}{l}{$\textit{,}\textit{ }$} & \multicolumn{6}{l}{$\emph{valueConstraint}\textit{ }$} & \multicolumn{2}{l}{$\textit{::}$} & \multicolumn{1}{l}{$\textit{ }$} & \multicolumn{3}{l}{$\textsc{UnionType}$} & \multicolumn{1}{l}{$\textit{ }$} & \multicolumn{1}{l}{$\textit{\}}\textit{}$}\\
\multicolumn{19}{l}{$\textbf{instance}\textit{ }\textsc{TypeCost}\textit{ }\textsc{MappingConstraint}$} & \multicolumn{1}{l}{$\textit{ }$} & \multicolumn{1}{l}{$\textbf{where}\textit{
}$}\\
\multicolumn{2}{p{2ex}}{  } & \multicolumn{6}{l}{$\emph{typeCost}$} & \multicolumn{1}{l}{$\textit{ }$} & \multicolumn{9}{l}{$\textsc{MappingNever}\textit{           }$} & \multicolumn{1}{l}{$\textit{=}$} & \multicolumn{1}{l}{$\textit{ }$} & \multicolumn{1}{l}{$0\textit{
}$}\\
\multicolumn{2}{p{2ex}}{  } & \multicolumn{6}{l}{$\emph{typeCost}$} & \multicolumn{1}{l}{$\textit{ }$} & \multicolumn{8}{l}{$\textsc{MappingConstraint}$} & \multicolumn{1}{l}{$\textit{ }\textit{\{}\textit{..}\textit{\}}\textit{ }$} & \multicolumn{1}{l}{$\textit{=}$} & \multicolumn{2}{l}{$\textit{
}$}\\
\multicolumn{6}{p{6ex}}{      } & \multicolumn{4}{l}{$\emph{typeCost}$} & \multicolumn{1}{l}{$\textit{ }$} & \multicolumn{6}{l}{$\emph{keyConstraint}$} & \multicolumn{4}{l}{$\textit{
}$}\\
\multicolumn{4}{p{4ex}}{    } & \multicolumn{2}{l}{$\textit{+}\textit{ }$} & \multicolumn{4}{l}{$\emph{typeCost}$} & \multicolumn{1}{l}{$\textit{ }$} & \multicolumn{10}{l}{$\emph{valueConstraint}$}\\

\end{tabular}

Separately, we acquire the information
about a possible typing of a JSON object
as a record of values. Note that
\texttt{RCTop} never actually occurs
during inference. That is, we could have
represented the
\texttt{RecordConstraint} as a
\texttt{Typelike} with an empty
\texttt{beyond} set. The merging of
constraints would be simply merging of
all column constraints.

\setlength{\tabcolsep}{1pt}
\begin{tabular}{llllllllllllllllll}
\multicolumn{4}{l}{$\textbf{data}$} & \multicolumn{3}{l}{$\textit{ }\textsc{RecordConstraint}\textit{ }$} & \multicolumn{1}{l}{$\textit{=}\textit{
}$}\\
\multicolumn{4}{p{4ex}}{    } & \multicolumn{2}{l}{$\textsc{RCTop}\textit{    }$} & \multicolumn{1}{l}{$\textit{-- beyond}$} & \multicolumn{1}{l}{$\textit{
}$}\\
\multicolumn{2}{p{2ex}}{  } & \multicolumn{1}{l}{$\textit{|}$} & \multicolumn{1}{l}{$\textit{ }$} & \multicolumn{2}{l}{$\textsc{RCBottom}\textit{ }$} & \multicolumn{1}{l}{$\textit{-- mempty}$} & \multicolumn{1}{l}{$\textit{
}$}\\
\multicolumn{2}{p{2ex}}{  } & \multicolumn{1}{l}{$\textit{|}$} & \multicolumn{1}{l}{$\textit{ }$} & \multicolumn{3}{l}{$\textsc{RecordConstraint}\textit{ }\textit{\{}$} & \multicolumn{1}{l}{$\textit{
}$}\\
\multicolumn{5}{p{6ex}}{      } & \multicolumn{3}{l}{$\emph{fields}\textit{ }\textit{::}\textit{ }\textsc{HashMap}\textit{ }\textsc{Text}\textit{ }\textsc{UnionType}\textit{ }\textit{\}}$}\\

\end{tabular}

\setlength{\tabcolsep}{1pt}
\begin{tabular}{lllllllllllllllllllllllllllll}
\multicolumn{4}{l}{$\textbf{instance}\textit{ }$} & \multicolumn{10}{l}{$\textsc{RecordConstraint}\textit{ }\textit{`}\textsc{Types}$} & \multicolumn{1}{l}{$\textit{`}$} & \multicolumn{1}{l}{$\textit{ }$} & \multicolumn{2}{l}{$\textsc{Object}\textit{ }\textbf{where}$} & \multicolumn{1}{l}{$\textit{
}$}\\
\multicolumn{2}{p{4ex}}{    } & \multicolumn{2}{l}{$\emph{infer}$} & \multicolumn{1}{l}{$\textit{ }$} & \multicolumn{1}{l}{$\textit{=}$} & \multicolumn{1}{l}{$\textit{ }$} & \multicolumn{7}{l}{$\textsc{RecordConstraint}\textit{    }$} & \multicolumn{1}{l}{$\textit{.}$} & \multicolumn{1}{l}{$\textit{ }$} & \multicolumn{2}{l}{$\emph{Map.fromList}$} & \multicolumn{1}{l}{$\textit{
}$}\\
\multicolumn{5}{p{10ex}}{          } & \multicolumn{1}{l}{$\textit{.}$} & \multicolumn{1}{l}{$\textit{ }$} & \multicolumn{7}{l}{$\emph{fmap}\textit{ }\textit{(}\emph{second}\textit{ }\emph{infer}\textit{)}\textit{ }$} & \multicolumn{1}{l}{$\textit{.}$} & \multicolumn{1}{l}{$\textit{ }$} & \multicolumn{3}{l}{$\emph{Map.toList}\textit{
}$}\\
\multicolumn{2}{p{4ex}}{    } & \multicolumn{2}{l}{$\emph{check}$} & \multicolumn{1}{l}{$\textit{ }$} & \multicolumn{11}{l}{$\textsc{RecordConstraint}\textit{ }\textit{\{}\emph{fields}$} & \multicolumn{1}{l}{$\textit{\}}\textit{ }$} & \multicolumn{2}{l}{$\emph{obj}\textit{ }\textit{=}\textit{
}$}\\
\multicolumn{4}{p{9ex}}{         } & \multicolumn{4}{l}{$\emph{all}\textit{ }\textit{(}\textit{`}\emph{elem}\textit{`}$} & \multicolumn{1}{l}{$\textit{ }$} & \multicolumn{4}{l}{$\emph{Map.keys}$} & \multicolumn{4}{l}{$\textit{ }\emph{fields}$} & \multicolumn{2}{l}{$\textit{)}\textit{
}$}\\
\multicolumn{8}{p{20ex}}{                    } & \multicolumn{1}{l}{$\textit{(}$} & \multicolumn{4}{l}{$\emph{Map.keys}$} & \multicolumn{6}{l}{$\textit{  }\emph{obj}\textit{)}\textit{
}$}\\
\multicolumn{3}{p{6ex}}{      } & \multicolumn{16}{l}{$\textit{\&\&}\textit{ }\emph{and}\textit{ }\textit{(}\emph{Map.elems}\textit{ }\textit{\$}\textit{ }\emph{Map.intersectionWith}\textit{
}$}\\
\multicolumn{12}{p{28ex}}{                            } & \multicolumn{4}{l}{$\emph{check}\textit{ }$} & \multicolumn{3}{l}{$\emph{fields}\textit{ }\emph{obj}\textit{)}\textit{
}$}\\
\multicolumn{3}{p{6ex}}{      } & \multicolumn{1}{l}{$\textit{\&\&}\textit{ }$} & \multicolumn{6}{l}{$\emph{all}\textit{ }\emph{isNullable}\textit{ }$} & \multicolumn{1}{l}{$\textit{(}$} & \multicolumn{5}{l}{$\emph{Map.elems}$} & \multicolumn{3}{l}{$\textit{
}$}\\
\multicolumn{10}{p{24ex}}{                        } & \multicolumn{1}{l}{$\textit{\$}$} & \multicolumn{5}{l}{$\textit{ }\emph{fields}\textit{ }\textit{`}$} & \multicolumn{3}{l}{$\emph{Map.difference}\textit{`}\textit{ }\emph{obj}\textit{)}\textit{
}$}\\
\multicolumn{4}{l}{$\textit{         }$} & \multicolumn{15}{l}{$\textit{-- absent values are nullable}$}\\

\end{tabular}

Observing that the two abstract domains
considered above are independent, we can
store the information about both options
separately in a record\footnote{The
  choice of representation will be
  explained later. Here we only consider
  acquiring information about possible
  values.}. It should be noted that this
representation is similar to
\emph{intersection type}: any value that
satisfies \texttt{ObjectConstraint} must
conform to both \texttt{mappingCase},
and \texttt{recordCase}. Also, this
\emph{intersection approach} in order to
address alternative union type
representations benefit from
\emph{principal type property}, meaning
that a principal type serves to acquire
the information corresponding to
different representations and handle
them separately. Since we plan to choose
only one representation for the object,
we can say that the minimum cost of this
type is the minimum of component costs.

\setlength{\tabcolsep}{1pt}
\begin{tabular}{lllllllllllllllllllll}
\multicolumn{3}{l}{$\textbf{data}\textit{ }$} & \multicolumn{3}{l}{$\textsc{ObjectConstraint}$} & \multicolumn{1}{l}{$\textit{ }$} & \multicolumn{1}{l}{$\textit{=}$} & \multicolumn{3}{l}{$\textit{ }\textsc{ObjectNever}\textit{ }\textit{-- mempty}\textit{
}$}\\
\multicolumn{2}{p{2ex}}{  } & \multicolumn{1}{l}{$\textit{|}\textit{  }$} & \multicolumn{3}{l}{$\textsc{ObjectConstraint}$} & \multicolumn{1}{l}{$\textit{ }$} & \multicolumn{1}{l}{$\textit{\{}$} & \multicolumn{3}{l}{$\textit{
}$}\\
\multicolumn{5}{p{9ex}}{         } & \multicolumn{1}{l}{$\emph{mappingCase}\textit{ }$} & \multicolumn{2}{l}{$\textit{::}$} & \multicolumn{1}{l}{$\textit{ }$} & \multicolumn{1}{l}{$\textsc{MappingConstraint}$} & \multicolumn{1}{l}{$\textit{
}$}\\
\multicolumn{4}{p{7ex}}{       } & \multicolumn{1}{l}{$\textit{,}\textit{ }$} & \multicolumn{1}{l}{$\emph{recordCase}\textit{  }$} & \multicolumn{2}{l}{$\textit{::}$} & \multicolumn{1}{l}{$\textit{ }$} & \multicolumn{1}{l}{$\textsc{RecordConstraint}\textit{ }$} & \multicolumn{1}{l}{$\textit{\}}\textit{ }$}\\

\end{tabular}

\setlength{\tabcolsep}{1pt}
\begin{tabular}{lllllllllllllllllll}
\multicolumn{5}{l}{$\textbf{instance}\textit{ }\textsc{TypeCost}\textit{ }\textsc{ObjectConstraint}$} & \multicolumn{1}{l}{$\textit{ }$} & \multicolumn{3}{l}{$\textbf{where}\textit{
}$}\\
\multicolumn{2}{p{2ex}}{  } & \multicolumn{1}{l}{$\emph{typeCost}\textit{ }\textsc{ObjectConstraint}\textit{ }\textit{\{}$} & \multicolumn{1}{l}{$\textit{..}\textit{\}}\textit{ }$} & \multicolumn{1}{l}{$\textit{=}$} & \multicolumn{1}{l}{$\textit{ }$} & \multicolumn{1}{l}{$\emph{typeCost}$} & \multicolumn{1}{l}{$\textit{ }$} & \multicolumn{1}{l}{$\emph{mappingCase}\textit{
}$}\\
\multicolumn{3}{p{29ex}}{                             } & \multicolumn{1}{l}{$\textit{`}\emph{min}$} & \multicolumn{1}{l}{$\textit{`}$} & \multicolumn{1}{l}{$\textit{ }$} & \multicolumn{1}{l}{$\emph{typeCost}$} & \multicolumn{1}{l}{$\textit{ }$} & \multicolumn{1}{l}{$\emph{recordCase}$}\\

\end{tabular}

\hypertarget{array-constraint}{%
\subsubsection{Array
constraint}\label{array-constraint}}

Similarly to the object type,
\texttt{ArrayConstraint} is used to
simultaneously obtain information about
all possible representations of an
array, differentiating between an array
of the same elements, and a row with the
type depending on a column. We need to
acquire the information for both
alternatives separately, and then, to
measure a relative likelihood of either
case, before mapping the union type to
Haskell declaration. Here, we specify
the records for two different possible
representations:

\setlength{\tabcolsep}{1pt}
\begin{tabular}{lllllllllllllllllllll}
\multicolumn{3}{l}{$\textbf{data}$} & \multicolumn{1}{l}{$\textit{ }$} & \multicolumn{1}{l}{$\textsc{ArrayConstraint}$} & \multicolumn{1}{l}{$\textit{ }$} & \multicolumn{1}{l}{$\textit{=}$} & \multicolumn{1}{l}{$\textit{ }$} & \multicolumn{1}{l}{$\textsc{ArrayNever}$} & \multicolumn{1}{l}{$\textit{ }$} & \multicolumn{1}{l}{$\textit{-- mempty}\textit{
}$}\\
\multicolumn{2}{p{3ex}}{   } & \multicolumn{1}{l}{$\textit{|}$} & \multicolumn{1}{l}{$\textit{ }$} & \multicolumn{1}{l}{$\textsc{ArrayConstraint}$} & \multicolumn{1}{l}{$\textit{ }$} & \multicolumn{1}{l}{$\textit{\{}$} & \multicolumn{1}{l}{$\textit{ }$} & \multicolumn{1}{l}{$\emph{rowCase}\textit{ }\textit{::}$} & \multicolumn{1}{l}{$\textit{ }$} & \multicolumn{1}{l}{$\textsc{RowConstraint}\textit{
}$}\\
\multicolumn{6}{p{21ex}}{                     } & \multicolumn{1}{l}{$\textit{,}$} & \multicolumn{1}{l}{$\textit{ }$} & \multicolumn{1}{l}{$\emph{arrayCase}\textit{ }$} & \multicolumn{2}{l}{$\textit{::}\textit{ }\textsc{UnionType}\textit{ }\textit{\}}$}\\

\end{tabular}

Semigroup operation just merges
information on the components, and the
same is done when inferring types or
checking them: For the arrays, we plan
to choose again only one of possible
representations, so the cost of
optionality is the lesser of the costs
of the representation-specific
constraints.

\setlength{\tabcolsep}{1pt}
\begin{tabular}{lllllllllllllllllllllllll}
\multicolumn{4}{l}{$\textbf{instance}\textit{ }$} & \multicolumn{6}{l}{$\textsc{ArrayConstraint}\textit{ }\textit{`}\textsc{Types}$} & \multicolumn{1}{l}{$\textit{`}$} & \multicolumn{1}{l}{$\textit{ }$} & \multicolumn{3}{l}{$\textsc{Array}\textit{ }\textbf{where}\textit{
}$}\\
\multicolumn{2}{p{4ex}}{    } & \multicolumn{2}{l}{$\emph{infer}$} & \multicolumn{6}{l}{$\textit{ }\emph{vs}\textit{ }\textit{=}\textit{ }\textsc{ArrayConstraint}\textit{ }$} & \multicolumn{1}{l}{$\textit{\{}$} & \multicolumn{1}{l}{$\textit{ }$} & \multicolumn{3}{l}{$\emph{rowCase}\textit{ }\textit{=}\textit{ }\emph{infer}\textit{ }\emph{vs}\textit{
}$}\\
\multicolumn{3}{p{6ex}}{      } & \multicolumn{5}{l}{$\textit{,}\textit{ }\emph{arrayCase}\textit{ }\textit{=}\textit{ }\emph{mconcat}\textit{ }\textit{(}$} & \multicolumn{7}{l}{$\emph{infer}\textit{ }\mathbin{\ooalign{\raise.29ex\hbox{$\scriptscriptstyle\$$}\cr\hss$\!\lozenge$\hss}}\textit{
}$}\\
\multicolumn{8}{p{29ex}}{                             } & \multicolumn{7}{l}{$\emph{Foldable.toList}\textit{ }\emph{vs}\textit{)}\textit{ }\textit{\}}\textit{
}$}\\
\multicolumn{2}{p{4ex}}{    } & \multicolumn{2}{l}{$\emph{check}$} & \multicolumn{1}{l}{$\textit{ }$} & \multicolumn{5}{l}{$\textsc{ArrayNever}\textit{           }$} & \multicolumn{2}{l}{$\emph{vs}$} & \multicolumn{1}{l}{$\textit{ }$} & \multicolumn{1}{l}{$\textit{=}$} & \multicolumn{1}{l}{$\textit{ }\textsc{False}\textit{
}$}\\
\multicolumn{2}{p{4ex}}{    } & \multicolumn{2}{l}{$\emph{check}$} & \multicolumn{1}{l}{$\textit{ }$} & \multicolumn{2}{l}{$\textsc{ArrayConstraint}$} & \multicolumn{3}{l}{$\textit{ }\textit{\{}\textit{..}\textit{\}}\textit{ }$} & \multicolumn{2}{l}{$\emph{vs}$} & \multicolumn{1}{l}{$\textit{ }$} & \multicolumn{1}{l}{$\textit{=}$} & \multicolumn{1}{l}{$\textit{
}$}\\
\multicolumn{4}{p{9ex}}{         } & \multicolumn{2}{l}{$\emph{check}$} & \multicolumn{1}{l}{$\textit{ }\emph{rowCase}\textit{ }\emph{vs}$} & \multicolumn{8}{l}{$\textit{
}$}\\
\multicolumn{3}{p{6ex}}{      } & \multicolumn{1}{l}{$\textit{\&\&}\textit{ }$} & \multicolumn{2}{l}{$\emph{and}\textit{ }\textit{(}$} & \multicolumn{2}{l}{$\emph{check}\textit{ }\emph{arrayCase}$} & \multicolumn{1}{l}{$\textit{ }$} & \multicolumn{6}{l}{$\mathbin{\ooalign{\raise.29ex\hbox{$\scriptscriptstyle\$$}\cr\hss$\!\lozenge$\hss}}\textit{
}$}\\
\multicolumn{6}{p{14ex}}{              } & \multicolumn{2}{l}{$\emph{Foldable.toList}$} & \multicolumn{1}{l}{$\textit{ }$} & \multicolumn{6}{l}{$\emph{vs}\textit{)}$}\\

\end{tabular}

\hypertarget{row-constraint}{%
\subsubsection{Row
constraint}\label{row-constraint}}

A row constraint is valid only if there
is the same number of entries in all
rows, which is represented by escaping
the \texttt{beyond} set whenever there
is an uneven number of columns. Row
constraint remains valid only if both
constraint describe the record of the
same length; otherwise, we yield
\texttt{RowTop} to indicate that it is
no longer valid. In other words,
\texttt{RowConstraint} is a
\emph{levitated
semilattice}{[}\protect\hyperlink{ref-levitated-lattice}{16}{]}\footnote{\emph{Levitated
  lattice} is created by appending
  distinct \texttt{bottom} and
  \texttt{top} to a set that does not
  possess them by itself.} with a
neutral element over the content type
that is a list of \texttt{UnionType}
objects.

\setlength{\tabcolsep}{1pt}
\begin{tabular}{lllllllllllllll}
\multicolumn{2}{l}{$\textbf{data}\textit{ }\textsc{RowConstraint}\textit{ }$} & \multicolumn{1}{l}{$\textit{=}$} & \multicolumn{1}{l}{$\textit{ }$} & \multicolumn{1}{l}{$\textsc{RowTop}\textit{ }\textit{|}\textit{ }\textsc{RowNever}\textit{
}$}\\
\multicolumn{2}{p{19ex}}{                   } & \multicolumn{1}{l}{$\textit{|}$} & \multicolumn{1}{l}{$\textit{ }$} & \multicolumn{1}{l}{$\textsc{Row}\textit{ }\textit{[}\textsc{UnionType}\textit{]}$}\\

\end{tabular}

\hypertarget{combining-the-union-type}{%
\subsubsection{Combining the union
type}\label{combining-the-union-type}}

It should note that given the
constraints for the different type
constructors, the union type can be
considered as mostly a generic
\texttt{Monoid} instance
{[}\protect\hyperlink{ref-generic-monoid}{11}{]}.
Merging information with
\texttt{\textless{}\textgreater{}} and
\texttt{mempty} follow the pattern
above, by just lifting operations on the
component.

\setlength{\tabcolsep}{1pt}
\begin{tabular}{llllllllllllllllllllll}
\multicolumn{4}{l}{$\textbf{data}$} & \multicolumn{3}{l}{$\textit{ }\textsc{UnionType}$} & \multicolumn{1}{l}{$\textit{ }\textit{=}$} & \multicolumn{1}{l}{$\textit{ }$} & \multicolumn{3}{l}{$\textsc{UnionType}\textit{ }\textit{\{}\textit{
}$}\\
\multicolumn{4}{p{4ex}}{    } & \multicolumn{2}{l}{$\emph{unionNull}$} & \multicolumn{1}{l}{$\textit{ }$} & \multicolumn{1}{l}{$\textit{::}$} & \multicolumn{1}{l}{$\textit{ }$} & \multicolumn{1}{l}{$\textsc{NullConstraint}$} & \multicolumn{2}{l}{$\textit{
}$}\\
\multicolumn{2}{p{2ex}}{  } & \multicolumn{1}{l}{$\textit{,}$} & \multicolumn{1}{l}{$\textit{ }$} & \multicolumn{2}{l}{$\emph{unionBool}$} & \multicolumn{1}{l}{$\textit{ }$} & \multicolumn{1}{l}{$\textit{::}$} & \multicolumn{1}{l}{$\textit{ }$} & \multicolumn{1}{l}{$\textsc{BoolConstraint}$} & \multicolumn{2}{l}{$\textit{
}$}\\
\multicolumn{2}{p{2ex}}{  } & \multicolumn{1}{l}{$\textit{,}$} & \multicolumn{1}{l}{$\textit{ }$} & \multicolumn{1}{l}{$\emph{unionNum}$} & \multicolumn{2}{l}{$\textit{  }$} & \multicolumn{1}{l}{$\textit{::}$} & \multicolumn{1}{l}{$\textit{ }$} & \multicolumn{2}{l}{$\textsc{NumberConstraint}$} & \multicolumn{1}{l}{$\textit{
}$}\\
\multicolumn{2}{p{2ex}}{  } & \multicolumn{1}{l}{$\textit{,}$} & \multicolumn{1}{l}{$\textit{ }$} & \multicolumn{1}{l}{$\emph{unionStr}$} & \multicolumn{2}{l}{$\textit{  }$} & \multicolumn{1}{l}{$\textit{::}$} & \multicolumn{1}{l}{$\textit{ }$} & \multicolumn{2}{l}{$\textsc{StringConstraint}$} & \multicolumn{1}{l}{$\textit{
}$}\\
\multicolumn{2}{p{2ex}}{  } & \multicolumn{1}{l}{$\textit{,}$} & \multicolumn{1}{l}{$\textit{ }$} & \multicolumn{1}{l}{$\emph{unionArr}$} & \multicolumn{2}{l}{$\textit{  }$} & \multicolumn{1}{l}{$\textit{::}$} & \multicolumn{1}{l}{$\textit{ }$} & \multicolumn{3}{l}{$\textsc{ArrayConstraint}\textit{
}$}\\
\multicolumn{2}{p{2ex}}{  } & \multicolumn{1}{l}{$\textit{,}$} & \multicolumn{1}{l}{$\textit{ }$} & \multicolumn{1}{l}{$\emph{unionObj}$} & \multicolumn{2}{l}{$\textit{  }$} & \multicolumn{1}{l}{$\textit{::}$} & \multicolumn{1}{l}{$\textit{ }$} & \multicolumn{3}{l}{$\textsc{ObjectConstraint}\textit{ }\textit{\}}$}\\

\end{tabular}

The generic structure of union type can
be explained by the fact that the
information contained in each record
field is \emph{independent} from the
information contained in other fields.
It means that we generalize
independently over different
dimensions\footnote{In this example,
  JSON terms can be described by terms
  without variables, and sets of tuples
  for dictionaries, so generalization by
  anti-unification is straightforward.}

Inference breaks down disjoint
alternatives corresponding to different
record fields, depending on the
constructor of a given value. It enables
implementing a clear and efficient
treatment of different alternatives
separately\footnote{The question may
  arise: what is the \emph{union type}
  without \emph{set union}? When the
  sets are disjoint, we just put the
  values in different bins to enable
  easier handling.}. Since union type is
all about optionality, we need to sum
all options from different alternatives
to obtain its \texttt{typeCost}.

\setlength{\tabcolsep}{1pt}
\begin{tabular}{llllllllllllllllllllllllllll}
\multicolumn{3}{l}{$\textbf{instance}$} & \multicolumn{1}{l}{$\textit{ }$} & \multicolumn{5}{l}{$\textsc{UnionType}$} & \multicolumn{1}{l}{$\textit{ }$} & \multicolumn{1}{l}{$\textit{`}$} & \multicolumn{1}{l}{$\textsc{Types}\textit{`}\textit{ }$} & \multicolumn{6}{l}{$\textsc{Value}\textit{ }\textbf{where}\textit{
}$}\\
\multicolumn{2}{p{2ex}}{  } & \multicolumn{1}{l}{$\emph{infer}\textit{ }$} & \multicolumn{1}{l}{$\textit{(}$} & \multicolumn{3}{l}{$\textsc{Bool}\textit{   }$} & \multicolumn{1}{l}{$\beta$} & \multicolumn{1}{l}{$\textit{)}$} & \multicolumn{1}{l}{$\textit{ }$} & \multicolumn{1}{l}{$\textit{=}$} & \multicolumn{1}{l}{$\textit{ }\emph{mempty}$} & \multicolumn{6}{l}{$\textit{ }\textit{\{}\textit{ }\emph{unionBool}\textit{ }\textit{=}\textit{
}$}\\
\multicolumn{4}{p{9ex}}{         } & \multicolumn{3}{l}{$\emph{infer}\textit{  }$} & \multicolumn{1}{l}{$\beta$} & \multicolumn{1}{l}{$\textit{ }$} & \multicolumn{1}{l}{$\textit{\}}$} & \multicolumn{8}{l}{$\textit{
}$}\\
\multicolumn{2}{p{2ex}}{  } & \multicolumn{2}{l}{$\emph{infer}\textit{  }$} & \multicolumn{6}{l}{$\textsc{Null}\textit{      }$} & \multicolumn{8}{l}{$\textit{=}\textit{ }\emph{mempty}\textit{ }\textit{\{}\textit{ }\emph{unionNull}\textit{ }\textit{=}\textit{
}$}\\
\multicolumn{4}{p{9ex}}{         } & \multicolumn{2}{l}{$\emph{infer}\textit{ }$} & \multicolumn{1}{l}{$\textit{(}$} & \multicolumn{1}{l}{$\textit{)}$} & \multicolumn{1}{l}{$\textit{ }$} & \multicolumn{1}{l}{$\textit{\}}$} & \multicolumn{8}{l}{$\textit{
}$}\\
\multicolumn{2}{p{2ex}}{  } & \multicolumn{2}{l}{$\emph{infer}\textit{ }\textit{(}$} & \multicolumn{2}{l}{$\textsc{Number}$} & \multicolumn{1}{l}{$\textit{ }$} & \multicolumn{1}{l}{$\emph{n}$} & \multicolumn{1}{l}{$\textit{)}$} & \multicolumn{1}{l}{$\textit{ }$} & \multicolumn{8}{l}{$\textit{=}\textit{ }\emph{mempty}\textit{ }\textit{\{}\textit{ }\emph{unionNum}\textit{  }\textit{=}\textit{
}$}\\
\multicolumn{4}{p{9ex}}{         } & \multicolumn{2}{l}{$\emph{infer}\textit{ }$} & \multicolumn{1}{l}{$\emph{n}$} & \multicolumn{2}{l}{$\textit{  }$} & \multicolumn{1}{l}{$\textit{\}}$} & \multicolumn{8}{l}{$\textit{
}$}\\
\multicolumn{2}{p{2ex}}{  } & \multicolumn{2}{l}{$\emph{infer}\textit{ }\textit{(}$} & \multicolumn{2}{l}{$\textsc{String}$} & \multicolumn{1}{l}{$\textit{ }$} & \multicolumn{2}{l}{$\emph{s}\textit{)}$} & \multicolumn{1}{l}{$\textit{ }$} & \multicolumn{8}{l}{$\textit{=}\textit{ }\emph{mempty}\textit{ }\textit{\{}\textit{ }\emph{unionStr}\textit{  }\textit{=}\textit{
}$}\\
\multicolumn{4}{p{9ex}}{         } & \multicolumn{2}{l}{$\emph{infer}\textit{ }$} & \multicolumn{1}{l}{$\emph{s}$} & \multicolumn{2}{l}{$\textit{  }$} & \multicolumn{1}{l}{$\textit{\}}$} & \multicolumn{8}{l}{$\textit{
}$}\\
\multicolumn{2}{p{2ex}}{  } & \multicolumn{2}{l}{$\emph{infer}\textit{ }\textit{(}$} & \multicolumn{2}{l}{$\textsc{Object}$} & \multicolumn{1}{l}{$\textit{ }$} & \multicolumn{2}{l}{$\emph{o}\textit{)}$} & \multicolumn{1}{l}{$\textit{ }$} & \multicolumn{8}{l}{$\textit{=}\textit{ }\emph{mempty}\textit{ }\textit{\{}\textit{ }\emph{unionObj}\textit{  }\textit{=}\textit{
}$}\\
\multicolumn{4}{p{9ex}}{         } & \multicolumn{1}{l}{$\emph{infer}$} & \multicolumn{1}{l}{$\textit{ }$} & \multicolumn{1}{l}{$\emph{o}$} & \multicolumn{2}{l}{$\textit{  }$} & \multicolumn{1}{l}{$\textit{\}}$} & \multicolumn{8}{l}{$\textit{
}$}\\
\multicolumn{2}{p{2ex}}{  } & \multicolumn{2}{l}{$\emph{infer}\textit{ }\textit{(}$} & \multicolumn{1}{l}{$\textsc{Array}$} & \multicolumn{2}{l}{$\textit{  }$} & \multicolumn{2}{l}{$\alpha\textit{)}$} & \multicolumn{1}{l}{$\textit{ }$} & \multicolumn{8}{l}{$\textit{=}\textit{ }\emph{mempty}\textit{ }\textit{\{}\textit{ }\emph{unionArr}\textit{  }\textit{=}\textit{
}$}\\
\multicolumn{4}{p{9ex}}{         } & \multicolumn{1}{l}{$\emph{infer}$} & \multicolumn{2}{l}{$\textit{ }\alpha$} & \multicolumn{2}{l}{$\textit{  }$} & \multicolumn{1}{l}{$\textit{\}}$} & \multicolumn{8}{l}{$\textit{
}$}\\
\multicolumn{18}{l}{$\textit{
}$}\\
\multicolumn{2}{p{2ex}}{  } & \multicolumn{1}{l}{$\emph{check}\textit{ }$} & \multicolumn{8}{l}{$\textsc{UnionType}\textit{ }\textit{\{}\textit{ }$} & \multicolumn{3}{l}{$\emph{unionBool}$} & \multicolumn{2}{l}{$\textit{ }\textit{\}}\textit{ }\textit{(}\textsc{Bool}\textit{   }$} & \multicolumn{1}{l}{$\beta$} & \multicolumn{1}{l}{$\textit{)}\textit{ }\textit{=}\textit{
}$}\\
\multicolumn{3}{p{8ex}}{        } & \multicolumn{8}{l}{$\emph{check}\textit{       }$} & \multicolumn{3}{l}{$\emph{unionBool}$} & \multicolumn{2}{l}{$\textit{           }$} & \multicolumn{1}{l}{$\beta$} & \multicolumn{1}{l}{$\textit{
}$}\\
\multicolumn{2}{p{2ex}}{  } & \multicolumn{1}{l}{$\emph{check}\textit{ }$} & \multicolumn{8}{l}{$\textsc{UnionType}\textit{ }\textit{\{}\textit{ }$} & \multicolumn{3}{l}{$\emph{unionNull}$} & \multicolumn{1}{l}{$\textit{ }\textit{\}}\textit{  }$} & \multicolumn{3}{l}{$\textsc{Null}\textit{      }\textit{=}\textit{
}$}\\
\multicolumn{3}{p{8ex}}{        } & \multicolumn{8}{l}{$\emph{check}\textit{       }$} & \multicolumn{3}{l}{$\emph{unionNull}$} & \multicolumn{1}{l}{$\textit{    }$} & \multicolumn{3}{l}{$\textit{(}\textit{)}\textit{
}$}\\
\multicolumn{2}{p{2ex}}{  } & \multicolumn{1}{l}{$\emph{check}\textit{ }$} & \multicolumn{8}{l}{$\textsc{UnionType}\textit{ }\textit{\{}\textit{ }$} & \multicolumn{2}{l}{$\emph{unionNum}$} & \multicolumn{2}{l}{$\textit{  }\textit{\}}\textit{ }\textit{(}$} & \multicolumn{1}{l}{$\textsc{Number}\textit{ }$} & \multicolumn{1}{l}{$\emph{n}$} & \multicolumn{1}{l}{$\textit{)}\textit{ }\textit{=}\textit{
}$}\\
\multicolumn{3}{p{8ex}}{        } & \multicolumn{8}{l}{$\emph{check}\textit{       }$} & \multicolumn{2}{l}{$\emph{unionNum}$} & \multicolumn{3}{l}{$\textit{            }$} & \multicolumn{1}{l}{$\emph{n}$} & \multicolumn{1}{l}{$\textit{
}$}\\
\multicolumn{2}{p{2ex}}{  } & \multicolumn{1}{l}{$\emph{check}\textit{ }$} & \multicolumn{8}{l}{$\textsc{UnionType}\textit{ }\textit{\{}\textit{ }$} & \multicolumn{2}{l}{$\emph{unionStr}$} & \multicolumn{3}{l}{$\textit{  }\textit{\}}\textit{ }\textit{(}\textsc{String}\textit{ }$} & \multicolumn{1}{l}{$\emph{s}$} & \multicolumn{1}{l}{$\textit{)}\textit{ }\textit{=}\textit{
}$}\\
\multicolumn{3}{p{8ex}}{        } & \multicolumn{8}{l}{$\emph{check}\textit{       }$} & \multicolumn{2}{l}{$\emph{unionStr}$} & \multicolumn{3}{l}{$\textit{            }$} & \multicolumn{1}{l}{$\emph{s}$} & \multicolumn{1}{l}{$\textit{
}$}\\
\multicolumn{2}{p{2ex}}{  } & \multicolumn{1}{l}{$\emph{check}\textit{ }$} & \multicolumn{8}{l}{$\textsc{UnionType}\textit{ }\textit{\{}\textit{ }$} & \multicolumn{2}{l}{$\emph{unionObj}$} & \multicolumn{3}{l}{$\textit{  }\textit{\}}\textit{ }\textit{(}\textsc{Object}\textit{ }$} & \multicolumn{1}{l}{$\emph{o}$} & \multicolumn{1}{l}{$\textit{)}\textit{ }\textit{=}\textit{
}$}\\
\multicolumn{3}{p{8ex}}{        } & \multicolumn{8}{l}{$\emph{check}\textit{       }$} & \multicolumn{2}{l}{$\emph{unionObj}$} & \multicolumn{3}{l}{$\textit{            }$} & \multicolumn{1}{l}{$\emph{o}$} & \multicolumn{1}{l}{$\textit{
}$}\\
\multicolumn{2}{p{2ex}}{  } & \multicolumn{1}{l}{$\emph{check}\textit{ }$} & \multicolumn{6}{l}{$\textsc{UnionType}\textit{ }$} & \multicolumn{2}{l}{$\textit{\{}\textit{ }$} & \multicolumn{2}{l}{$\emph{unionArr}$} & \multicolumn{3}{l}{$\textit{  }\textit{\}}\textit{ }\textit{(}\textsc{Array}\textit{  }$} & \multicolumn{1}{l}{$\alpha$} & \multicolumn{1}{l}{$\textit{)}\textit{ }\textit{=}\textit{
}$}\\
\multicolumn{3}{p{8ex}}{        } & \multicolumn{6}{l}{$\emph{check}\textit{     }$} & \multicolumn{7}{l}{$\emph{unionArr}\textit{              }$} & \multicolumn{2}{l}{$\alpha$}\\

\end{tabular}

\hypertarget{overlapping-alternatives}{%
\subsubsection{Overlapping
alternatives}\label{overlapping-alternatives}}

The essence of union type systems have
long been dealing with the conflicting
types provided in the input. Motivated
by the examples above, we also aim to
address conflicting alternative
assignments. It is apparent that
examples 4. to 6. hint at more than one
assignment: in example 5, a set of lists
of values that may correspond to
\texttt{Int}, \texttt{String}, or
\texttt{null}, or a table that has the
same (and predefined) type for each
values; in example 6 A record of fixed
names or the mapping from hash to a
single object type.

\hypertarget{counting-observations}{%
\subsubsection{Counting
observations}\label{counting-observations}}

In this section, we discuss how to
gather information about the number of
samples supporting each alternative type
constraint. To explain this, the other
example can be considered:

\begin{Shaded}
\begin{Highlighting}[]
\FunctionTok{\{}\DataTypeTok{"history"}\FunctionTok{:} \OtherTok{[}
   \FunctionTok{\{}\DataTypeTok{"error"}  \FunctionTok{:} \StringTok{"Authorization failed"}\FunctionTok{,}
    \DataTypeTok{"code"}   \FunctionTok{:}  \DecValTok{401}\FunctionTok{\}}
  \OtherTok{,}\FunctionTok{\{}\DataTypeTok{"message"}\FunctionTok{:} \StringTok{"Where can I submit my proposal?"}\FunctionTok{,}
    \DataTypeTok{"uid"}    \FunctionTok{:} \DecValTok{1014}\FunctionTok{\}}
  \OtherTok{,}\FunctionTok{\{}\DataTypeTok{"message"}\FunctionTok{:} \StringTok{"Sent it to HotCRP"}\FunctionTok{,}
    \DataTypeTok{"uid"}    \FunctionTok{:} \DecValTok{93}\FunctionTok{\}}
  \OtherTok{,}\FunctionTok{\{}\DataTypeTok{"message"}\FunctionTok{:} \StringTok{"Thanks!"}\FunctionTok{,}
    \DataTypeTok{"uid"}    \FunctionTok{:} \DecValTok{1014}\FunctionTok{\}}
  \OtherTok{,}\FunctionTok{\{}\DataTypeTok{"error"}  \FunctionTok{:} \StringTok{"Authorization failed"}\FunctionTok{,}
    \DataTypeTok{"code"}   \FunctionTok{:}  \DecValTok{401}\FunctionTok{\}}\OtherTok{]}\FunctionTok{\}}
\end{Highlighting}
\end{Shaded}

First, we need to identify it as a list
of similar elements. Second, there are
multiple instances of each record
example. We consider that the best
approach would be to use the multisets
of inferred records instead. To find the
best representation, we can a type
complexity, and attempt to minimize the
term. Next step is to detect the
similarities between type descriptions
introduced for different parts of the
term:

\begin{Shaded}
\begin{Highlighting}[]
\FunctionTok{\{}\DataTypeTok{"history"}      \FunctionTok{:} \OtherTok{[}\ErrorTok{...}\OtherTok{]}
\FunctionTok{,}\DataTypeTok{"last\_message"} \FunctionTok{:} \FunctionTok{\{}\DataTypeTok{"message"}\FunctionTok{:} \StringTok{"Thanks!"}\FunctionTok{,}
                   \DataTypeTok{"uid"} \FunctionTok{:} \DecValTok{1014}\FunctionTok{\}} \FunctionTok{\}}
\end{Highlighting}
\end{Shaded}

We can add the auxiliary information
about a number of samples observed, and
the constraint will remain a
\texttt{Typelike} object. The
\texttt{Counted} constraint counts the
number of samples observed for the
constraint inside so that we can decide
on which alternative representation is
best supported by evidence. It should be
noted that \texttt{Counted} constraint
is the first example that does not
correspond to a semilattice, that is
\texttt{a\textless{}\textgreater{}a}\(\neq\)\texttt{a}.
This is natural for a \texttt{Typelike}
object; it is not a type constraint in a
conventional sense, just an accumulation
of knowledge.

\setlength{\tabcolsep}{1pt}
\begin{tabular}{llllllllllll}
\multicolumn{1}{l}{$\textbf{data}\textit{ }\textsc{Counted}\textit{ }\alpha\textit{ }\textit{=}\textit{ }\textsc{Counted}\textit{ }\textit{\{}\textit{ }\emph{count}\textit{::}\textsc{Int}\textit{,}\textit{ }\emph{constraint}\textit{::}\alpha\textit{ }\textit{\}}$}\\

\end{tabular}

\setlength{\tabcolsep}{1pt}
\begin{tabular}{llllllllllllllllllllllllllllll}
\multicolumn{4}{l}{$\textbf{instance}$} & \multicolumn{1}{l}{$\textit{ }$} & \multicolumn{3}{l}{$\textsc{Semigroup}$} & \multicolumn{1}{l}{$\textit{ }$} & \multicolumn{1}{l}{$\alpha$} & \multicolumn{10}{l}{$\textit{
}$}\\
\multicolumn{3}{p{6ex}}{      } & \multicolumn{1}{l}{$\textit{=>}$} & \multicolumn{1}{l}{$\textit{ }$} & \multicolumn{3}{l}{$\textsc{Semigroup}$} & \multicolumn{1}{l}{$\textit{ }$} & \multicolumn{1}{l}{$\textit{(}$} & \multicolumn{10}{l}{$\textsc{Counted}\textit{ }\alpha\textit{)}\textit{ }\textbf{where}\textit{
}$}\\
\multicolumn{2}{p{2ex}}{  } & \multicolumn{1}{l}{$\alpha\textit{ }\diamond$} & \multicolumn{1}{l}{$\textit{ }\beta$} & \multicolumn{1}{l}{$\textit{ }$} & \multicolumn{3}{l}{$\textit{=}\textit{ }\textsc{Counted}$} & \multicolumn{1}{l}{$\textit{ }$} & \multicolumn{1}{l}{$\textit{\{}$} & \multicolumn{10}{l}{$\textit{
}$}\\
\multicolumn{7}{p{14ex}}{              } & \multicolumn{2}{l}{$\emph{count}$} & \multicolumn{2}{l}{$\textit{      }$} & \multicolumn{1}{l}{$\textit{=}$} & \multicolumn{1}{l}{$\textit{ }$} & \multicolumn{1}{l}{$\emph{count}\textit{      }$} & \multicolumn{1}{l}{$\alpha$} & \multicolumn{1}{l}{$\textit{ }$} & \multicolumn{1}{c}{$\textit{+}\textit{  }$} & \multicolumn{1}{l}{$\emph{count}\textit{      }$} & \multicolumn{1}{l}{$\beta$} & \multicolumn{1}{l}{$\textit{
}$}\\
\multicolumn{6}{p{12ex}}{            } & \multicolumn{1}{l}{$\textit{,}\textit{ }$} & \multicolumn{4}{l}{$\emph{constraint}\textit{ }$} & \multicolumn{1}{l}{$\textit{=}$} & \multicolumn{1}{l}{$\textit{ }$} & \multicolumn{1}{l}{$\emph{constraint}\textit{ }$} & \multicolumn{1}{l}{$\alpha$} & \multicolumn{1}{l}{$\textit{ }$} & \multicolumn{1}{c}{$\diamond\textit{ }$} & \multicolumn{1}{l}{$\emph{constraint}\textit{ }$} & \multicolumn{1}{l}{$\beta$} & \multicolumn{1}{l}{$\textit{
}$}\\
\multicolumn{6}{p{12ex}}{            } & \multicolumn{14}{l}{$\textit{\}}$}\\

\end{tabular}

Therefore, at each step, we may need to
maintain a \textbf{cardinality} of each
possible value, and is provided with
sufficient number of samples, we may
attempt to detect\footnote{If we detect
  a pattern too early, we risk to make
  the types too narrow to work with
  actual API responses.}. To preserve
efficiency, we may need to merge
whenever the number of alternatives in a
multiset crosses the threshold. We can
attempt to narrow strings only in the
cases when cardinality crosses the
threshold.

\hypertarget{finishing-touches}{%
\section{Finishing
touches}\label{finishing-touches}}

The final touch would be to perform the
post-processing of an assigned type
before generating it to make it more
resilient to common uncertainties. These
assumptions may bypass the defined
least-upper-bound criterion specified in
the initial part of the paper; however,
they prove to work well in
practice{[}\protect\hyperlink{ref-quicktype}{2},
\protect\hyperlink{ref-json-autotype-prezi}{14}{]}.

If we have no observations corresponding
to an array type, it can be inconvenient
to disallow an array to contain any
values at all. Therefore, we introduce a
non-monotonic step of converting the
\texttt{mempty} into a final
\texttt{Typelike} object aiming to
introduce a representation allowing the
occurrence of any \texttt{Value} in the
input. That still preserves the validity
of the typing. We note that the program
using our types must not have any
assumptions about these values; however,
at the same time, it should be able to
print them for debugging purposes.

In most JSON documents, we observe that
the same object can be simultaneously
described in different parts of sample
data structures. Due to this reason, we
compare the sets of labels assigned to
all objects and propose to unify those
that have more than 60\% of identical
labels. For transparency, the identified
candidates are logged for each user, and
a user can also indicate them explicitly
instead of relying on automation. We
conclude that this allows considerably
decreasing the complexity of types and
makes the output less redundant.

\hypertarget{future-work}{%
\section{Future
work}\label{future-work}}

In the present paper, we only discuss
typing of tree-like values. However, it
is natural to scale this approach to
multiple types in APIs, in which
different types are referred to by name
and possibly contain each other. To
address these cases, we plan to show
that the environment of
\texttt{Typelike} objects is also
\texttt{Typelike}, and that constraint
generalization (\emph{anti-unification})
can be extended in the same way.

It should be noted that many
\texttt{Typelike} instances for
non-simple types usually follow one the
two patterns of (1) for a finite sum of
disjoint constructors, we bin this
information by each constructor during
the inference (2) for typing terms with
multiple alternative representations, we
infer all constraints separately for
each alternative representation. In both
cases, \texttt{Generic} derivation
procedure for the \texttt{Monoid},
\texttt{Typelike}, and \texttt{TypeCost}
instances is possible
{[}\protect\hyperlink{ref-generics}{17}{]}.
This allows us to design a type system
by declaring datatypes themselves and
leave implementation to the compiler.
Manual implementation would be only left
for special cases, like
\texttt{StringConstraint} and
\texttt{Counted} constraint.

Finally, we believe that we can explain
the duality of categorical framework of
\texttt{Typelike} categories and use
generalization (anti-unification)
instead of unification (or narrowing) as
a type inference mechanism. The
\texttt{beyond} set would then
correspond to a set of error messages,
and a result of the inference would
represent a principal type in
Damas-Milner sense.

\hypertarget{conclusion}{%
\subsection{Conclusion}\label{conclusion}}

In the present study, we aimed to derive
the types that were valid with respect
to the provided
specification\footnote{Specification was
  given in the motivation section
  descriptions of JSON input examples,
  and the expected results given as
  Haskell type declarations.}, thereby
obtaining the information from the input
in the most comprehensive way. We
defined type inference as representation
learning and type system engineering as
a meta-learning problem in which the
\textbf{priors corresponding to the data
structure induced typing rules}. We show
how the type safety can be quickly
tested as equational laws with
QuickCheck, which is a useful
prototyping tool, and may be
supplemented with fully formal proof in
the future.

We also formulated the \textbf{union
type discipline} as manipulation of
\texttt{Typelike} commutative monoids,
that represented knowledge about the
data structure. In addition, we proposed
a union type system engineering
methodology that was logically justified
by theoretical criteria. We demonstrated
that it was capable of consistently
explaining the decisions made in
practice. We followed a strictly
constructive procedure, that can be
implemented generically.

We hope that this kind of
straightforward type system engineering
will become widely used in practice,
replacing less modular approaches of the
past. The proposed approach may be used
to underlie the way towards formal
construction and derivation of type
systems based on the specification of
value domains and design constraints.

\hypertarget{bibliography}{%
\section*{Bibliography}\label{bibliography}}
\addcontentsline{toc}{section}{Bibliography}

\hypertarget{refs}{}
\begin{cslreferences}
\leavevmode\hypertarget{ref-aeson}{}%
{[}1{]} Aeson: Fast JSON parsing and
generation: 2011.
\emph{\url{https://hackage.haskell.org/package/aeson}}.

\leavevmode\hypertarget{ref-quicktype}{}%
{[}2{]} A first look at quicktype: 2017.
\emph{\url{https://blog.quicktype.io/first-look/}}.

\leavevmode\hypertarget{ref-javascript-inference}{}%
{[}3{]} Anderson, C. et al. 2005.
Towards type inference for javascript.
\emph{ECOOP 2005 - object-oriented
programming} (Berlin, Heidelberg, 2005),
428--452.

\leavevmode\hypertarget{ref-quickcheck}{}%
{[}4{]} Claessen, K. and Hughes, J.
2000. QuickCheck: A lightweight tool for
random testing of haskell programs.
\emph{SIGPLAN Not.} 35, 9 (Sep. 2000),
268--279.
DOI:\url{https://doi.org/10.1145/357766.351266}.

\leavevmode\hypertarget{ref-undefined1}{}%
{[}5{]} C Standard undefined behaviour
versus Wittgenstein:
\emph{\url{https://www.yodaiken.com/2018/05/20/depressing-and-faintly-terrifying-days-for-the-c-standard/}}.

\leavevmode\hypertarget{ref-undefined2}{}%
{[}6{]} C Undefined Behavior -
Depressing and Terrifying (Updated):
2018.
\emph{\url{https://www.yodaiken.com/2018/05/20/depressing-and-faintly-terrifying-days-for-the-c-standard/}}.

\leavevmode\hypertarget{ref-entangled}{}%
{[}7{]} EnTangleD: A bi-directional
literate programming tool: 2019.
\emph{\url{https://blog.esciencecenter.nl/entangled-1744448f4b9f}}.

\leavevmode\hypertarget{ref-pads}{}%
{[}8{]} Fisher, K. and Walker, D. 2011.
The PADS Project: An Overview.
\emph{Proceedings of the 14th
international conference on database
theory} (New York, NY, USA, 2011),
11--17.

\leavevmode\hypertarget{ref-semantic-subtyping}{}%
{[}9{]} Frisch, A. et al. 2002. Semantic
subtyping. \emph{Proceedings 17th annual
ieee symposium on logic in computer
science} (2002), 137--146.

\leavevmode\hypertarget{ref-semantic-subtyping2}{}%
{[}10{]} Frisch, A. et al. 2008.
Semantic subtyping: Dealing
set-theoretically with function, union,
intersection, and negation types.
\emph{J. ACM}. 55, 4 (Sep. 2008).
DOI:\url{https://doi.org/10.1145/1391289.1391293}.

\leavevmode\hypertarget{ref-generic-monoid}{}%
{[}11{]} Generics example: Creating
monoid instances: 2012.
\emph{\url{https://www.yesodweb.com/blog/2012/10/generic-monoid}}.

\leavevmode\hypertarget{ref-ghcid}{}%
{[}12{]} GHCID - a new ghci based ide
(ish): 2014.
\emph{\url{http://neilmitchell.blogspot.com/2014/09/ghcid-new-ghci-based-ide-ish.html}}.

\leavevmode\hypertarget{ref-xduce}{}%
{[}13{]} Hosoya, H. and Pierce, B. 2000.
XDuce: A typed xml processing language.
(Jun. 2000).

\leavevmode\hypertarget{ref-json-autotype-prezi}{}%
{[}14{]} JSON autotype: Presentation for
Haskell.SG: 2015.
\emph{\url{https://engineers.sg/video/json-autotype-1-0-haskell-sg--429}}.

\leavevmode\hypertarget{ref-literate-programming}{}%
{[}15{]} Knuth, D.E. 1984. Literate
programming. \emph{Comput. J.} 27, 2
(May 1984), 97--111.
DOI:\url{https://doi.org/10.1093/comjnl/27.2.97}.

\leavevmode\hypertarget{ref-levitated-lattice}{}%
{[}16{]} Lattices: Fine-grained library
for constructing and manipulating
lattices: 2017.
\emph{\url{http://hackage.haskell.org/package/lattices-2.0.2/docs/Algebra-Lattice-Levitated.html}}.

\leavevmode\hypertarget{ref-generics}{}%
{[}17{]} Magalhães, J.P. et al. 2010. A
generic deriving mechanism for haskell.
\emph{SIGPLAN Not.} 45, 11 (Sep. 2010),
37--48.
DOI:\url{https://doi.org/10.1145/2088456.1863529}.

\leavevmode\hypertarget{ref-xml-typelift}{}%
{[}18{]} Michal J. Gajda, D.K. 2020.
Fast XML/HTML tools for Haskell: XML
Typelift and improved Xeno. Manuscript
under review.

\leavevmode\hypertarget{ref-pandoc}{}%
{[}19{]} Pandoc: A universal document
converter: 2000.
\emph{\url{https://pandoc.org}}.

\leavevmode\hypertarget{ref-type-providers-f-sharp}{}%
{[}20{]} Petricek, T. et al. 2016. Types
from Data: Making Structured Data
First-Class Citizens in F\#.
\emph{SIGPLAN Not.} 51, 6 (Jun. 2016),
477--490.
DOI:\url{https://doi.org/10.1145/2980983.2908115}.

\leavevmode\hypertarget{ref-GHCZurihac}{}%
{[}21{]} Peyton Jones, S. 2019. Type
inference as constraint solving: How
ghc's type inference engine actually
works. Zurihac keynote talk.

\leavevmode\hypertarget{ref-gradual-typing}{}%
{[}22{]} Siek, J. and Taha, W. 2007.
Gradual typing for objects.
\emph{Proceedings of the 21st european
conference on object-oriented
programming} (Berlin, Heidelberg, 2007),
2--27.

\leavevmode\hypertarget{ref-HM-X}{}%
{[}23{]} Sulzmann, M. and Stuckey, P.j.
2008. Hm(x) Type Inference is Clp(x)
Solving. \emph{J. Funct. Program.} 18, 2
(Mar. 2008), 251--283.
DOI:\url{https://doi.org/10.1017/S0956796807006569}.

\leavevmode\hypertarget{ref-subtype-inequalities}{}%
{[}24{]} Tiuryn, J. 1992. Subtype
inequalities. \emph{{[}1992{]}
Proceedings of the Seventh Annual IEEE
Symposium on Logic in Computer Science}.
(1992), 308--315.

\leavevmode\hypertarget{ref-subtyping-lattice}{}%
{[}25{]} Tiuryn, J. 1997. Subtyping over
a lattice (abstract).
\emph{Computational logic and proof
theory} (Berlin, Heidelberg, 1997),
84--88.

\leavevmode\hypertarget{ref-undefined3}{}%
{[}26{]} Undefined behavior in 2017:
2017.
\emph{\url{https://blog.regehr.org/archives/1520}}.

\leavevmode\hypertarget{ref-typescript-soundness}{}%
{[}27{]} 2019.
\emph{\url{https://github.com/microsoft/TypeScript/issues/9825}}.
\end{cslreferences}

\hypertarget{appendix-all-laws-of-typelike}{%
\section{Appendix: all laws of
Typelike}\label{appendix-all-laws-of-typelike}}

\[
\begin{array}{l l l l lllllcr}
                    &     &   &             & \textrm{check} & \textrm{mempty} & v & = & \textbf{False} & \ & \textit{(mempty contains no terms)} \\
    \textrm{beyond} & t   &   & \Rightarrow & \textrm{check} & t               & v & = & \textbf{True} & \hspace{1cm}\hfill{} & \textit{(beyond contains all terms)} \\
    \textrm{check}  & t_1 & v & \Rightarrow & \textrm{check} & (t_1 \diamond t_2) & v & = & \textbf{True} & \hspace{1cm}\hfill{} & \textit{(left fusion keeps terms)} \\
    \textrm{check}  & t_2 & v & \Rightarrow & \textrm{check} & (t_1 \diamond t_2) & v & = & \textbf{True} & \hspace{1cm}\hfill{} & \textit{(right fusion keeps terms)} \\
                    &     &   &             & \textrm{check} & (\textrm{infer}\ v) & v & = & \textbf{False} & \ & \textit{(inferred type contains the source term)} \\
                    &     &   &             \multicolumn{4}{r}{t_1 \diamond (t_2 \diamond t_3)} & = & t_1 \diamond (t_2 \diamond t_3) & \ & \textit{(semigroup associativity)} \\
                    &     &   &             \multicolumn{4}{r}{\textrm{mempty} \diamond t} & = & t & \ & \textit{(left identity of the monoid)} \\
                    &     &   &             \multicolumn{4}{r}{t \diamond \textrm{mempty}} & = & t & \ & \textit{(right identity of the monoid)} \\
    \end{array} \]

\hypertarget{appendix-module-headers}{%
\section{Appendix: definition module
headers}\label{appendix-module-headers}}

\setlength{\tabcolsep}{1pt}
\begin{tabular}{lllllllllllllllllllllllllllllllllll}
\multicolumn{22}{l}{$\textit{\{-\# language AllowAmbiguousTypes        \#-\}}$} & \multicolumn{3}{l}{$\textit{}$}\\
\multicolumn{22}{l}{$\textit{\{-\# language DeriveGeneric              \#-\}}$} & \multicolumn{3}{l}{$\textit{}$}\\
\multicolumn{22}{l}{$\textit{\{-\# language DuplicateRecordFields      \#-\}}$} & \multicolumn{3}{l}{$\textit{}$}\\
\multicolumn{22}{l}{$\textit{\{-\# language FlexibleInstances          \#-\}}$} & \multicolumn{3}{l}{$\textit{}$}\\
\multicolumn{22}{l}{$\textit{\{-\# language GeneralizedNewtypeDeriving \#-\}}$} & \multicolumn{3}{l}{$\textit{}$}\\
\multicolumn{22}{l}{$\textit{\{-\# language MultiParamTypeClasses      \#-\}}$} & \multicolumn{3}{l}{$\textit{}$}\\
\multicolumn{22}{l}{$\textit{\{-\# language NamedFieldPuns             \#-\}}$} & \multicolumn{3}{l}{$\textit{}$}\\
\multicolumn{22}{l}{$\textit{\{-\# language PartialTypeSignatures      \#-\}}$} & \multicolumn{3}{l}{$\textit{}$}\\
\end{tabular}

\hypertarget{sec:test-suite}{%
\section{Appendix: test
suite}\label{sec:test-suite}}

\setlength{\tabcolsep}{1pt}
\begin{tabular}{llllllllllllllllllllllllllllllllllllllllllllllllllllllllllllllll}
\multicolumn{43}{l}{$\textit{\{-\# language FlexibleInstances     \#-\}}$} & \multicolumn{18}{l}{$\textit{}$}\\
\multicolumn{43}{l}{$\textit{\{-\# language Rank2Types            \#-\}}$} & \multicolumn{18}{l}{$\textit{}$}\\
\multicolumn{43}{l}{$\textit{\{-\# language MultiParamTypeClasses \#-\}}$} & \multicolumn{18}{l}{$\textit{}$}\\
\multicolumn{43}{l}{$\textit{\{-\# language MultiWayIf            \#-\}}$} & \multicolumn{18}{l}{$\textit{}$}\\
\multicolumn{43}{l}{$\textit{\{-\# language NamedFieldPuns        \#-\}}$} & \multicolumn{18}{l}{$\textit{}$}\\
\multicolumn{43}{l}{$\textit{\{-\# language ScopedTypeVariables   \#-\}}$} & \multicolumn{18}{l}{$\textit{}$}\\
\multicolumn{43}{l}{$\textit{\{-\# language StandaloneDeriving    \#-\}}$} & \multicolumn{18}{l}{$\textit{}$}\\
\multicolumn{43}{l}{$\textit{\{-\# language TemplateHaskell       \#-\}}$} & \multicolumn{18}{l}{$\textit{}$}\\
\end{tabular}

\hypertarget{appendix-package-dependencies}{%
\section{Appendix: package
dependencies}\label{appendix-package-dependencies}}

\begin{Shaded}
\begin{Highlighting}[]
\FunctionTok{name}\KeywordTok{:}\AttributeTok{ union{-}types}
\FunctionTok{version}\KeywordTok{:}\AttributeTok{ }\StringTok{\textquotesingle{}0.1.0.0\textquotesingle{}}
\FunctionTok{category}\KeywordTok{:}\AttributeTok{ Web}
\FunctionTok{author}\KeywordTok{:}\AttributeTok{ Anonymous}
\FunctionTok{maintainer}\KeywordTok{:}\AttributeTok{ example@example.com}
\FunctionTok{license}\KeywordTok{:}\AttributeTok{ BSD{-}3}
\FunctionTok{extra{-}source{-}files}\KeywordTok{:}
\KeywordTok{{-}}\AttributeTok{ CHANGELOG.md}
\KeywordTok{{-}}\AttributeTok{ README.md}
\FunctionTok{dependencies}\KeywordTok{:}
\KeywordTok{{-}}\AttributeTok{ base}
\KeywordTok{{-}}\AttributeTok{ aeson}
\KeywordTok{{-}}\AttributeTok{ containers}
\KeywordTok{{-}}\AttributeTok{ text}
\KeywordTok{{-}}\AttributeTok{ hspec}
\KeywordTok{{-}}\AttributeTok{ QuickCheck}
\KeywordTok{{-}}\AttributeTok{ unordered{-}containers}
\KeywordTok{{-}}\AttributeTok{ scientific}
\KeywordTok{{-}}\AttributeTok{ hspec}
\KeywordTok{{-}}\AttributeTok{ QuickCheck}
\KeywordTok{{-}}\AttributeTok{ validity}
\KeywordTok{{-}}\AttributeTok{ vector}
\KeywordTok{{-}}\AttributeTok{ unordered{-}containers}
\KeywordTok{{-}}\AttributeTok{ scientific}
\KeywordTok{{-}}\AttributeTok{ genvalidity}
\KeywordTok{{-}}\AttributeTok{ genvalidity{-}hspec}
\KeywordTok{{-}}\AttributeTok{ genvalidity{-}property}
\KeywordTok{{-}}\AttributeTok{ time}
\KeywordTok{{-}}\AttributeTok{ email{-}validate}
\KeywordTok{{-}}\AttributeTok{ generic{-}arbitrary}
\KeywordTok{{-}}\AttributeTok{ mtl}
\KeywordTok{{-}}\AttributeTok{ hashable}
\FunctionTok{library}\KeywordTok{:}
\AttributeTok{  }\FunctionTok{source{-}dirs}\KeywordTok{:}\AttributeTok{ src}
\AttributeTok{  }\FunctionTok{exposed{-}modules}\KeywordTok{:}
\AttributeTok{  }\KeywordTok{{-}}\AttributeTok{ Unions}
\FunctionTok{tests}\KeywordTok{:}
\AttributeTok{  }\FunctionTok{spec}\KeywordTok{:}
\AttributeTok{    }\FunctionTok{main}\KeywordTok{:}\AttributeTok{ Spec.hs}
\AttributeTok{    }\FunctionTok{source{-}dirs}\KeywordTok{:}
\AttributeTok{      }\KeywordTok{{-}}\AttributeTok{ test/spec}
\AttributeTok{    }\FunctionTok{dependencies}\KeywordTok{:}
\AttributeTok{      }\KeywordTok{{-}}\AttributeTok{ union{-}types}
\AttributeTok{      }\KeywordTok{{-}}\AttributeTok{ mtl}
\AttributeTok{      }\KeywordTok{{-}}\AttributeTok{ random}
\AttributeTok{      }\KeywordTok{{-}}\AttributeTok{ transformers}
\AttributeTok{      }\KeywordTok{{-}}\AttributeTok{ hashable}
\AttributeTok{      }\KeywordTok{{-}}\AttributeTok{ quickcheck{-}classes}
\AttributeTok{      }\KeywordTok{{-}}\AttributeTok{ file{-}embed}
\AttributeTok{      }\KeywordTok{{-}}\AttributeTok{ bytestring}
\AttributeTok{      }\KeywordTok{{-}}\AttributeTok{ less{-}arbitrary}
\end{Highlighting}
\end{Shaded}

\hypertarget{appendix-representation-of-generated-haskell-types}{%
\section{Appendix: representation of
generated Haskell
types}\label{appendix-representation-of-generated-haskell-types}}

We will not delve here into identifier
conversion between JSON and Haskell, so
it suffices that we have an abstract
datatypes for Haskell type and
constructor identifiers:

\setlength{\tabcolsep}{1pt}
\begin{tabular}{llllllllllllllll}
\multicolumn{3}{l}{$\textbf{newtype}\textit{ }\textsc{HConsId}$} & \multicolumn{1}{l}{$\textit{  }\textit{=}$} & \multicolumn{1}{l}{$\textit{ }$} & \multicolumn{1}{l}{$\textsc{HConsId}\textit{ }\textsc{String}\textit{
}$}\\
\multicolumn{2}{p{2ex}}{  } & \multicolumn{1}{l}{$\textbf{deriving}\textit{ }\textit{(}\textsc{Eq}\textit{,}$} & \multicolumn{1}{l}{$\textsc{Ord}$} & \multicolumn{1}{l}{$\textit{,}$} & \multicolumn{1}{l}{$\textsc{Show}\textit{,}\textsc{Generic}\textit{,}\textsc{IsString}\textit{)}\textit{}$}\\
\multicolumn{4}{l}{$\textbf{newtype}\textit{ }\textsc{HFieldId}\textit{ }\textit{=}$} & \multicolumn{1}{l}{$\textit{ }$} & \multicolumn{1}{l}{$\textsc{HFieldId}\textit{ }\textsc{String}\textit{
}$}\\
\multicolumn{2}{p{2ex}}{  } & \multicolumn{1}{l}{$\textbf{deriving}\textit{ }\textit{(}\textsc{Eq}\textit{,}$} & \multicolumn{1}{l}{$\textsc{Ord}$} & \multicolumn{1}{l}{$\textit{,}$} & \multicolumn{1}{l}{$\textsc{Show}\textit{,}\textsc{Generic}\textit{,}\textsc{IsString}\textit{)}\textit{}$}\\
\multicolumn{3}{l}{$\textbf{newtype}\textit{ }\textsc{HTypeId}$} & \multicolumn{1}{l}{$\textit{  }\textit{=}$} & \multicolumn{1}{l}{$\textit{ }$} & \multicolumn{1}{l}{$\textsc{HTypeId}\textit{ }\textsc{String}\textit{
}$}\\
\multicolumn{2}{p{2ex}}{  } & \multicolumn{1}{l}{$\textbf{deriving}\textit{ }\textit{(}\textsc{Eq}\textit{,}$} & \multicolumn{1}{l}{$\textsc{Ord}$} & \multicolumn{1}{l}{$\textit{,}$} & \multicolumn{1}{l}{$\textsc{Show}\textit{,}\textsc{Generic}\textit{,}\textsc{IsString}\textit{)}$}\\

\end{tabular}

For each single type we will either
describe its exact representation or
reference to the other definition by
name:

\setlength{\tabcolsep}{1pt}
\begin{tabular}{llllllllllllllllllll}
\multicolumn{4}{l}{$\textbf{data}$} & \multicolumn{6}{l}{$\textit{ }\textsc{HType}\textit{ }\textit{=}\textit{
}$}\\
\multicolumn{4}{p{4ex}}{    } & \multicolumn{1}{l}{$\textsc{HRef}$} & \multicolumn{1}{l}{$\textit{ }$} & \multicolumn{3}{l}{$\textsc{HTypeId}$} & \multicolumn{1}{l}{$\textit{
}$}\\
\multicolumn{2}{p{2ex}}{  } & \multicolumn{1}{l}{$\textit{|}$} & \multicolumn{1}{l}{$\textit{ }$} & \multicolumn{1}{l}{$\textsc{HApp}$} & \multicolumn{1}{l}{$\textit{ }$} & \multicolumn{3}{l}{$\textsc{HTypeId}$} & \multicolumn{1}{l}{$\textit{ }\textit{[}\textsc{HType}\textit{]}\textit{
}$}\\
\multicolumn{2}{p{2ex}}{  } & \multicolumn{1}{l}{$\textit{|}$} & \multicolumn{1}{l}{$\textit{ }$} & \multicolumn{1}{l}{$\textsc{HADT}$} & \multicolumn{1}{l}{$\textit{ }$} & \multicolumn{1}{l}{$\textit{[}$} & \multicolumn{1}{l}{$\textsc{HCons}$} & \multicolumn{1}{l}{$\textit{]}$} & \multicolumn{1}{l}{$\textit{
}$}\\
\multicolumn{2}{p{2ex}}{  } & \multicolumn{5}{l}{$\textbf{deriving}$} & \multicolumn{1}{l}{$\textit{ }\textit{(}\textsc{Eq}\textit{,}$} & \multicolumn{1}{l}{$\textit{ }$} & \multicolumn{1}{l}{$\textsc{Ord}\textit{,}\textit{ }\textsc{Show}\textit{,}\textit{ }\textsc{Generic}\textit{)}$}\\

\end{tabular}

For syntactic convenience, we will allow
string literals to denote type
references:

\setlength{\tabcolsep}{1pt}
\begin{tabular}{lllllllllllll}
\multicolumn{3}{l}{$\textbf{instance}\textit{ }\textsc{IsString}\textit{ }\textsc{HType}\textit{ }\textbf{where}\textit{
}$}\\
\multicolumn{2}{p{2ex}}{  } & \multicolumn{1}{l}{$\emph{fromString}\textit{ }\textit{=}\textit{ }\textsc{HRef}\textit{ }\textit{.}\textit{ }\emph{fromString}$}\\

\end{tabular}

When we define a single constructor, we
allow field and constructor names to be
empty strings (\texttt{""}), assuming
that the relevant identifiers will be
put there by post-processing that will
pick names using types of fields and
their containers
{[}\protect\hyperlink{ref-xml-typelift}{18}{]}.

\setlength{\tabcolsep}{1pt}
\begin{tabular}{llllllllllllllllllll}
\multicolumn{6}{l}{$\textbf{data}\textit{ }\textsc{HCons}\textit{ }\textit{=}\textit{ }\textsc{HCons}$} & \multicolumn{1}{l}{$\textit{ }$} & \multicolumn{3}{l}{$\textit{\{}\textit{
}$}\\
\multicolumn{5}{p{14ex}}{              } & \multicolumn{1}{l}{$\emph{name}$} & \multicolumn{1}{l}{$\textit{ }$} & \multicolumn{1}{l}{$\textit{::}$} & \multicolumn{1}{l}{$\textit{  }$} & \multicolumn{1}{l}{$\textsc{HConsId}\textit{
}$}\\
\multicolumn{3}{p{12ex}}{            } & \multicolumn{1}{l}{$\textit{,}$} & \multicolumn{1}{l}{$\textit{ }$} & \multicolumn{1}{l}{$\emph{args}$} & \multicolumn{1}{l}{$\textit{ }$} & \multicolumn{1}{l}{$\textit{::}$} & \multicolumn{1}{l}{$\textit{ }\textit{[}$} & \multicolumn{1}{l}{$\textit{(}\textsc{HFieldId}\textit{,}\textit{ }\textsc{HType}\textit{)}\textit{]}\textit{
}$}\\
\multicolumn{3}{p{12ex}}{            } & \multicolumn{1}{l}{$\textit{\}}$} & \multicolumn{6}{l}{$\textit{
}$}\\
\multicolumn{2}{p{2ex}}{  } & \multicolumn{1}{l}{$\textbf{deriving}\textit{ }\textit{(}$} & \multicolumn{7}{l}{$\textsc{Eq}\textit{,}\textit{ }\textsc{Ord}\textit{,}\textit{ }\textsc{Show}\textit{,}\textit{ }\textsc{Generic}\textit{)}$}\\

\end{tabular}

At some stage we want to split
representation into individually named
declarations, and then we use
environment of defined types, with an
explicitly named toplevel type:

\setlength{\tabcolsep}{1pt}
\begin{tabular}{llllllllllllllllll}
\multicolumn{3}{l}{$\textbf{data}$} & \multicolumn{1}{l}{$\textit{ }\textsc{HTypeEnv}$} & \multicolumn{1}{l}{$\textit{ }\textit{=}$} & \multicolumn{1}{l}{$\textit{ }$} & \multicolumn{2}{l}{$\textsc{HTypeEnv}\textit{ }\textit{\{}\textit{
}$}\\
\multicolumn{3}{p{4ex}}{    } & \multicolumn{1}{l}{$\emph{toplevel}\textit{ }$} & \multicolumn{1}{l}{$\textit{::}$} & \multicolumn{1}{l}{$\textit{ }$} & \multicolumn{1}{l}{$\textsc{HTypeId}$} & \multicolumn{1}{l}{$\textit{
}$}\\
\multicolumn{2}{p{2ex}}{  } & \multicolumn{1}{l}{$\textit{,}\textit{ }$} & \multicolumn{1}{l}{$\emph{env}\textit{      }$} & \multicolumn{1}{l}{$\textit{::}$} & \multicolumn{1}{l}{$\textit{ }$} & \multicolumn{1}{l}{$\textsc{HashMap}$} & \multicolumn{1}{l}{$\textit{ }\textsc{HTypeId}\textit{ }\textsc{HType}\textit{
}$}\\
\multicolumn{2}{p{2ex}}{  } & \multicolumn{6}{l}{$\textit{\}}$}\\

\end{tabular}

When checking for validity of types and
type environments, we might need a list
of predefined identifiers that are
imported:

\setlength{\tabcolsep}{1pt}
\begin{tabular}{lllllllllllllllllllll}
\multicolumn{7}{l}{$\emph{predefinedHTypes}$} & \multicolumn{1}{l}{$\textit{ }$} & \multicolumn{1}{l}{$\textit{::}$} & \multicolumn{1}{l}{$\textit{ }$} & \multicolumn{1}{l}{$\textit{[}\textsc{HType}\textit{]}\textit{}$}\\
\multicolumn{7}{l}{$\emph{predefinedHTypes}$} & \multicolumn{1}{l}{$\textit{ }$} & \multicolumn{1}{l}{$\textit{=}\textit{ }$} & \multicolumn{1}{l}{$\textit{[}$} & \multicolumn{1}{l}{$\textit{
}$}\\
\multicolumn{4}{p{4ex}}{    } & \multicolumn{7}{l}{$\textit{"Data.Aeson.Value"}\textit{
}$}\\
\multicolumn{2}{p{2ex}}{  } & \multicolumn{1}{l}{$\textit{,}$} & \multicolumn{1}{l}{$\textit{ }$} & \multicolumn{7}{l}{$\textit{"()"}\textit{
}$}\\
\multicolumn{2}{p{2ex}}{  } & \multicolumn{1}{l}{$\textit{,}$} & \multicolumn{1}{l}{$\textit{ }$} & \multicolumn{2}{l}{$\textit{"Double"}$} & \multicolumn{5}{l}{$\textit{
}$}\\
\multicolumn{2}{p{2ex}}{  } & \multicolumn{1}{l}{$\textit{,}$} & \multicolumn{1}{l}{$\textit{ }$} & \multicolumn{2}{l}{$\textit{"String"}$} & \multicolumn{5}{l}{$\textit{
}$}\\
\multicolumn{2}{p{2ex}}{  } & \multicolumn{1}{l}{$\textit{,}$} & \multicolumn{1}{l}{$\textit{ }$} & \multicolumn{7}{l}{$\textit{"Int"}\textit{
}$}\\
\multicolumn{2}{p{2ex}}{  } & \multicolumn{1}{l}{$\textit{,}$} & \multicolumn{1}{l}{$\textit{ }$} & \multicolumn{1}{l}{$\textit{"Date"}\textit{ }$} & \multicolumn{6}{l}{$\textit{-- actually: "Data.Time.CalendarDay"}\textit{
}$}\\
\multicolumn{2}{p{2ex}}{  } & \multicolumn{1}{l}{$\textit{,}$} & \multicolumn{1}{l}{$\textit{ }$} & \multicolumn{1}{l}{$\textit{"Email"}$} & \multicolumn{6}{l}{$\textit{ }\textit{-- actually: "Data.Email"}\textit{
}$}\\
\multicolumn{2}{p{2ex}}{  } & \multicolumn{9}{l}{$\textit{]}$}\\

\end{tabular}

Consider that we also have an
\texttt{htop} value that represents any
possible JSON value. It is polimorphic
for ease of use:

\setlength{\tabcolsep}{1pt}
\begin{tabular}{lllllllllllllll}
\multicolumn{1}{l}{$\emph{htop}$} & \multicolumn{1}{l}{$\textit{ }$} & \multicolumn{1}{l}{$\textit{::}$} & \multicolumn{1}{l}{$\textit{ }\textsc{IsString}\textit{ }\emph{s}\textit{ }\textit{=>}\textit{ }\emph{s}\textit{}$}\\
\multicolumn{1}{l}{$\emph{htop}$} & \multicolumn{1}{l}{$\textit{ }$} & \multicolumn{1}{l}{$\textit{=}\textit{ }$} & \multicolumn{1}{l}{$\textit{"Data.Aeson.Value"}$}\\

\end{tabular}

\hypertarget{code-for-selecting-representation}{%
\subsection{Code for selecting
representation}\label{code-for-selecting-representation}}

Below is the code to select Haskell type
representation. To convert union type
discipline to strict Haskell type
representations, we need to join the
options to get the actual
representation:

\setlength{\tabcolsep}{1pt}
\begin{tabular}{llllllllllllllllllllllllll}
\multicolumn{4}{l}{$\emph{toHType}$} & \multicolumn{3}{l}{$\textit{ }\textit{::}$} & \multicolumn{1}{l}{$\textit{ }$} & \multicolumn{4}{l}{$\textsc{ToHType}\textit{ }$} & \multicolumn{1}{l}{$\emph{ty}$} & \multicolumn{1}{l}{$\textit{ }$} & \multicolumn{1}{l}{$\textit{=>}\textit{ }\emph{ty}\textit{ }\textit{->}$} & \multicolumn{1}{l}{$\textit{ }\textsc{HType}\textit{}$}\\
\multicolumn{4}{l}{$\emph{toHType}$} & \multicolumn{3}{l}{$\textit{  }\textit{=}$} & \multicolumn{1}{l}{$\textit{ }$} & \multicolumn{4}{l}{$\emph{joinAlts}$} & \multicolumn{1}{l}{$\textit{ }\textit{.}$} & \multicolumn{1}{l}{$\textit{ }$} & \multicolumn{1}{l}{$\emph{toHTypes}$} & \multicolumn{1}{l}{$\textit{
}$}\\
\multicolumn{16}{l}{$\textit{}$}\\
\multicolumn{5}{l}{$\emph{joinAlts}$} & \multicolumn{5}{l}{$\textit{     }\textit{::}$} & \multicolumn{1}{l}{$\textit{ }\textit{[}$} & \multicolumn{3}{l}{$\textsc{HType}$} & \multicolumn{2}{l}{$\textit{]}\textit{ }\textit{->}\textit{ }\textsc{HType}\textit{}$}\\
\multicolumn{5}{l}{$\emph{joinAlts}$} & \multicolumn{1}{l}{$\textit{ }$} & \multicolumn{3}{l}{$\textit{[}\textit{]}\textit{   }$} & \multicolumn{1}{l}{$\textit{=}$} & \multicolumn{1}{l}{$\textit{  }$} & \multicolumn{3}{l}{$\emph{htop}\textit{ }$} & \multicolumn{2}{l}{$\textit{-- promotion of empty type}\textit{}$}\\
\multicolumn{5}{l}{$\emph{joinAlts}$} & \multicolumn{1}{l}{$\textit{ }$} & \multicolumn{3}{l}{$\emph{alts}\textit{ }$} & \multicolumn{1}{l}{$\textit{=}$} & \multicolumn{1}{l}{$\textit{  }$} & \multicolumn{5}{l}{$\emph{foldr1}\textit{ }\emph{joinPair}\textit{ }\emph{alts}\textit{
}$}\\
\multicolumn{2}{p{2ex}}{  } & \multicolumn{14}{l}{$\textbf{where}\textit{
}$}\\
\multicolumn{3}{p{4ex}}{    } & \multicolumn{13}{l}{$\emph{joinPair}\textit{ }\alpha\textit{ }\beta\textit{ }\textit{=}\textit{ }\textsc{HApp}\textit{ }\textit{":|:"}\textit{ }\textit{[}\alpha\textit{,}\textit{ }\beta\textit{]}$}\\

\end{tabular}

Considering the assembly of
\texttt{UnionType}, we join all the
options, and convert nullable types to
\texttt{Maybe} types

\setlength{\tabcolsep}{1pt}
\begin{tabular}{llllllllllllllllllllllllllllllll}
\multicolumn{11}{l}{$\textbf{instance}\textit{ }\textsc{ToHType}\textit{ }\textsc{UnionType}$} & \multicolumn{1}{l}{$\textit{ }$} & \multicolumn{10}{l}{$\textbf{where}\textit{
}$}\\
\multicolumn{2}{p{2ex}}{  } & \multicolumn{6}{l}{$\emph{toHTypes}\textit{ }\textsc{UnionType}\textit{ }$} & \multicolumn{1}{l}{$\textit{\{}$} & \multicolumn{2}{l}{$\textit{..}\textit{\}}\textit{ }$} & \multicolumn{1}{l}{$\textit{=}$} & \multicolumn{10}{l}{$\textit{
}$}\\
\multicolumn{4}{p{6ex}}{      } & \multicolumn{4}{l}{$\emph{prependNullable}$} & \multicolumn{1}{l}{$\textit{ }$} & \multicolumn{13}{l}{$\emph{unionNull}\textit{ }\emph{opts}\textit{
}$}\\
\multicolumn{3}{p{4ex}}{    } & \multicolumn{19}{l}{$\textbf{where}\textit{
}$}\\
\multicolumn{4}{p{6ex}}{      } & \multicolumn{3}{l}{$\emph{opts}\textit{ }\textit{=}\textit{ }\emph{concat}\textit{ }$} & \multicolumn{1}{l}{$\textit{[}$} & \multicolumn{6}{l}{$\emph{toHTypes}$} & \multicolumn{1}{l}{$\textit{ }$} & \multicolumn{7}{l}{$\emph{unionBool}\textit{
}$}\\
\multicolumn{7}{p{20ex}}{                    } & \multicolumn{1}{l}{$\textit{,}$} & \multicolumn{6}{l}{$\emph{toHTypes}$} & \multicolumn{1}{l}{$\textit{ }$} & \multicolumn{1}{l}{$\emph{unionStr}$} & \multicolumn{6}{l}{$\textit{
}$}\\
\multicolumn{7}{p{20ex}}{                    } & \multicolumn{1}{l}{$\textit{,}$} & \multicolumn{6}{l}{$\emph{toHTypes}$} & \multicolumn{1}{l}{$\textit{ }$} & \multicolumn{1}{l}{$\emph{unionNum}$} & \multicolumn{6}{l}{$\textit{
}$}\\
\multicolumn{7}{p{20ex}}{                    } & \multicolumn{1}{l}{$\textit{,}$} & \multicolumn{6}{l}{$\emph{toHTypes}$} & \multicolumn{1}{l}{$\textit{ }$} & \multicolumn{1}{l}{$\emph{unionArr}$} & \multicolumn{6}{l}{$\textit{
}$}\\
\end{tabular}

The type class returns a list of
mutually exclusive type representations:

\setlength{\tabcolsep}{1pt}
\begin{tabular}{lllllllllllllllllll}
\multicolumn{4}{l}{$\textbf{class}$} & \multicolumn{1}{l}{$\textit{ }$} & \multicolumn{2}{l}{$\textsc{Typelike}\textit{ }$} & \multicolumn{1}{l}{$\emph{ty}$} & \multicolumn{1}{l}{$\textit{
}$}\\
\multicolumn{3}{p{3ex}}{   } & \multicolumn{1}{l}{$\textit{=>}$} & \multicolumn{1}{l}{$\textit{ }$} & \multicolumn{1}{l}{$\textsc{ToHType}$} & \multicolumn{1}{l}{$\textit{  }$} & \multicolumn{1}{l}{$\emph{ty}$} & \multicolumn{1}{l}{$\textit{ }\textbf{where}\textit{
}$}\\
\multicolumn{2}{p{2ex}}{  } & \multicolumn{4}{l}{$\emph{toHTypes}\textit{ }\textit{::}$} & \multicolumn{2}{l}{$\textit{ }\emph{ty}\textit{ }$} & \multicolumn{1}{l}{$\textit{->}\textit{ }\textit{[}\textsc{HType}\textit{]}$}\\

\end{tabular}

Conversion of flat types is quite
straightforward:

\setlength{\tabcolsep}{1pt}
\begin{tabular}{llllllllllllllllllllllllllllllllll}
\multicolumn{11}{l}{$\textbf{instance}\textit{ }\textsc{ToHType}\textit{ }$} & \multicolumn{13}{l}{$\textsc{BoolConstraint}\textit{ }\textbf{where}\textit{
}$}\\
\multicolumn{2}{p{2ex}}{  } & \multicolumn{3}{l}{$\emph{toHTypes}$} & \multicolumn{1}{l}{$\textit{ }$} & \multicolumn{5}{l}{$\textsc{Absent}$} & \multicolumn{2}{l}{$\textit{  }$} & \multicolumn{1}{l}{$\textit{=}$} & \multicolumn{1}{l}{$\textit{ }$} & \multicolumn{1}{l}{$\textit{[}$} & \multicolumn{8}{l}{$\textit{]}\textit{
}$}\\
\multicolumn{2}{p{2ex}}{  } & \multicolumn{3}{l}{$\emph{toHTypes}$} & \multicolumn{1}{l}{$\textit{ }$} & \multicolumn{7}{l}{$\textsc{Present}\textit{ }$} & \multicolumn{1}{l}{$\textit{=}$} & \multicolumn{1}{l}{$\textit{ }$} & \multicolumn{1}{l}{$\textit{[}$} & \multicolumn{8}{l}{$\textit{"Bool"}\textit{]}\textit{}$}\\
\multicolumn{24}{l}{$\textbf{instance}\textit{ }\textsc{ToHType}\textit{ }\textsc{NumberConstraint}\textit{ }\textbf{where}\textit{
}$}\\
\multicolumn{2}{p{2ex}}{  } & \multicolumn{3}{l}{$\emph{toHTypes}$} & \multicolumn{1}{l}{$\textit{ }$} & \multicolumn{6}{l}{$\textsc{NCNever}$} & \multicolumn{1}{l}{$\textit{ }$} & \multicolumn{1}{l}{$\textit{=}$} & \multicolumn{1}{l}{$\textit{ }$} & \multicolumn{1}{l}{$\textit{[}$} & \multicolumn{8}{l}{$\textit{]}\textit{
}$}\\
\multicolumn{2}{p{2ex}}{  } & \multicolumn{3}{l}{$\emph{toHTypes}$} & \multicolumn{1}{l}{$\textit{ }$} & \multicolumn{6}{l}{$\textsc{NCFloat}$} & \multicolumn{1}{l}{$\textit{ }$} & \multicolumn{1}{l}{$\textit{=}$} & \multicolumn{1}{l}{$\textit{ }$} & \multicolumn{1}{l}{$\textit{[}$} & \multicolumn{8}{l}{$\textit{"Double"}\textit{]}\textit{
}$}\\
\end{tabular}

For array and object types we pick the
representation which presents the lowest
cost of optionality:

\setlength{\tabcolsep}{1pt}
\begin{tabular}{lllllllllllllllllllllllllllllllllllll}
\multicolumn{25}{l}{$\textbf{instance}\textit{ }\textsc{ToHType}\textit{ }\textsc{ObjectConstraint}$} & \multicolumn{1}{l}{$\textit{ }$} & \multicolumn{1}{l}{$\textbf{where}\textit{
}$}\\
\multicolumn{2}{p{2ex}}{  } & \multicolumn{5}{l}{$\emph{toHTypes}$} & \multicolumn{1}{l}{$\textit{ }$} & \multicolumn{17}{l}{$\textsc{ObjectNever}\textit{           }$} & \multicolumn{1}{l}{$\textit{=}$} & \multicolumn{1}{l}{$\textit{ }\textit{[}\textit{]}\textit{
}$}\\
\multicolumn{2}{p{2ex}}{  } & \multicolumn{5}{l}{$\emph{toHTypes}$} & \multicolumn{1}{l}{$\textit{ }$} & \multicolumn{11}{l}{$\textsc{ObjectConstraint}$} & \multicolumn{2}{l}{$\textit{ }\textit{\{}$} & \multicolumn{4}{l}{$\textit{..}\textit{\}}\textit{ }$} & \multicolumn{1}{l}{$\textit{=}$} & \multicolumn{1}{l}{$\textit{
}$}\\
\multicolumn{3}{p{4ex}}{    } & \multicolumn{1}{l}{$\textbf{if}$} & \multicolumn{15}{l}{$\textit{ }\emph{typeCost}\textit{ }\emph{recordCase}\textit{ }$} & \multicolumn{2}{l}{$\leq$} & \multicolumn{1}{l}{$\textit{ }$} & \multicolumn{5}{l}{$\emph{typeCost}\textit{ }\emph{mappingCase}\textit{
}$}\\
\multicolumn{4}{p{6ex}}{      } & \multicolumn{3}{l}{$\textbf{then}$} & \multicolumn{1}{l}{$\textit{ }$} & \multicolumn{4}{l}{$\emph{toHTypes}$} & \multicolumn{1}{l}{$\textit{ }$} & \multicolumn{9}{l}{$\emph{recordCase}$} & \multicolumn{5}{l}{$\textit{
}$}\\
\multicolumn{4}{p{6ex}}{      } & \multicolumn{3}{l}{$\textbf{else}$} & \multicolumn{1}{l}{$\textit{ }$} & \multicolumn{4}{l}{$\emph{toHTypes}$} & \multicolumn{1}{l}{$\textit{ }$} & \multicolumn{14}{l}{$\emph{mappingCase}\textit{
}$}\\
\multicolumn{27}{l}{$\textit{}$}\\
\multicolumn{27}{l}{$\textbf{instance}\textit{ }\textsc{ToHType}\textit{ }\textsc{RecordConstraint}\textit{ }\textbf{where}\textit{
}$}\\
\end{tabular}

\hypertarget{appendix-missing-pieces-of-code}{%
\section*{Appendix: Missing pieces of
code}\label{appendix-missing-pieces-of-code}}
\addcontentsline{toc}{section}{Appendix:
Missing pieces of code}

In order to represent \texttt{FreeType}
for the \texttt{Value}, we need to add
\texttt{Ord} instance for it:

\setlength{\tabcolsep}{1pt}
\begin{tabular}{llllllllllll}
\multicolumn{1}{l}{$\textbf{deriving}\textit{ }\textbf{instance}\textit{ }\textsc{Ord}\textit{ }\textsc{Value}$}\\

\end{tabular}

For validation of dates and emails, we
import functions from Hackage:

\setlength{\tabcolsep}{1pt}
\begin{tabular}{lllllllllllllllllllllllllll}
\multicolumn{4}{l}{$\emph{isValidDate}$} & \multicolumn{1}{l}{$\textit{ }$} & \multicolumn{2}{l}{$\textit{::}$} & \multicolumn{4}{l}{$\textit{ }\textsc{Text}\textit{ }$} & \multicolumn{6}{l}{$\textit{->}\textit{ }\textsc{Bool}\textit{}$}\\
\multicolumn{4}{l}{$\emph{isValidDate}$} & \multicolumn{1}{l}{$\textit{ }$} & \multicolumn{1}{l}{$\textit{=}$} & \multicolumn{1}{l}{$\textit{ }$} & \multicolumn{4}{l}{$\emph{isJust}$} & \multicolumn{6}{l}{$\textit{
}$}\\
\multicolumn{5}{p{12ex}}{            } & \multicolumn{1}{l}{$\textit{.}$} & \multicolumn{1}{l}{$\textit{ }$} & \multicolumn{10}{l}{$\emph{parseDate}\textit{
}$}\\
\multicolumn{5}{p{12ex}}{            } & \multicolumn{1}{l}{$\textit{.}$} & \multicolumn{1}{l}{$\textit{ }$} & \multicolumn{10}{l}{$\emph{Text.unpack}\textit{
}$}\\
\multicolumn{2}{p{2ex}}{  } & \multicolumn{15}{l}{$\textbf{where}\textit{
}$}\\
\multicolumn{3}{p{4ex}}{    } & \multicolumn{3}{l}{$\emph{parseDate}$} & \multicolumn{3}{l}{$\textit{ }\textit{::}$} & \multicolumn{1}{l}{$\textit{ }$} & \multicolumn{2}{l}{$\textsc{String}\textit{ }\textit{->}\textit{ }$} & \multicolumn{2}{l}{$\textsc{Maybe}$} & \multicolumn{3}{l}{$\textit{ }\textsc{Day}\textit{
}$}\\
\multicolumn{3}{p{4ex}}{    } & \multicolumn{3}{l}{$\emph{parseDate}$} & \multicolumn{3}{l}{$\textit{  }\textit{=}$} & \multicolumn{1}{l}{$\textit{ }$} & \multicolumn{2}{l}{$\emph{parseTimeM}$} & \multicolumn{1}{l}{$\textit{ }$} & \multicolumn{1}{l}{$\textsc{True}$} & \multicolumn{3}{l}{$\textit{
}$}\\
\multicolumn{13}{p{28ex}}{                            } & \multicolumn{2}{l}{$\emph{defaultTimeLocale}$} & \multicolumn{1}{l}{$\textit{ }$} & \multicolumn{1}{l}{$\textit{\$}\textit{
}$}\\
\end{tabular}

\setlength{\tabcolsep}{1pt}
\begin{tabular}{llllllllllllllllllllllllll}
\multicolumn{5}{l}{$\textbf{instance}\textit{ }$} & \multicolumn{1}{l}{$\textit{(}$} & \multicolumn{2}{l}{$\textsc{Hashable}$} & \multicolumn{8}{l}{$\textit{          }\emph{k}\textit{
}$}\\
\multicolumn{5}{p{9ex}}{         } & \multicolumn{1}{l}{$\textit{,}$} & \multicolumn{2}{l}{$\textsc{Hashable}$} & \multicolumn{4}{l}{$\textit{            }$} & \multicolumn{1}{l}{$\emph{v}$} & \multicolumn{1}{l}{$\textit{)}$} & \multicolumn{2}{l}{$\textit{
}$}\\
\multicolumn{3}{p{6ex}}{      } & \multicolumn{3}{l}{$\textit{=>}\textit{  }$} & \multicolumn{2}{l}{$\textsc{Hashable}$} & \multicolumn{1}{l}{$\textit{ }$} & \multicolumn{3}{l}{$\textit{(}\textsc{HashMap}\textit{ }\emph{k}\textit{ }$} & \multicolumn{1}{l}{$\emph{v}$} & \multicolumn{1}{l}{$\textit{)}$} & \multicolumn{1}{l}{$\textit{ }$} & \multicolumn{1}{l}{$\textbf{where}\textit{
}$}\\
\multicolumn{2}{p{2ex}}{  } & \multicolumn{5}{l}{$\emph{hashWithSalt}\textit{ }\emph{s}\textit{ }$} & \multicolumn{1}{l}{$\textit{=}$} & \multicolumn{1}{l}{$\textit{ }$} & \multicolumn{4}{l}{$\emph{hashWithSalt}$} & \multicolumn{1}{l}{$\textit{ }$} & \multicolumn{1}{l}{$\emph{s}$} & \multicolumn{1}{l}{$\textit{
}$}\\
\multicolumn{7}{p{17ex}}{                 } & \multicolumn{1}{l}{$\textit{.}$} & \multicolumn{1}{l}{$\textit{ }$} & \multicolumn{7}{l}{$\emph{Foldable.toList}\textit{
}$}\\
\multicolumn{16}{l}{$\textit{}$}\\
\multicolumn{4}{l}{$\textbf{instance}$} & \multicolumn{1}{l}{$\textit{ }$} & \multicolumn{2}{l}{$\textsc{Hashable}$} & \multicolumn{3}{l}{$\textit{           }$} & \multicolumn{1}{l}{$\emph{v}$} & \multicolumn{5}{l}{$\textit{
}$}\\
\end{tabular}

Then we put all the missing code in the
module:

\setlength{\tabcolsep}{1pt}
\begin{tabular}{llllllllllllllllllllllllllllllll}
\multicolumn{18}{l}{$\textit{\{-\# language AllowAmbiguousTypes        \#-\}}$} & \multicolumn{3}{l}{$\textit{}$}\\
\multicolumn{18}{l}{$\textit{\{-\# language DeriveGeneric              \#-\}}$} & \multicolumn{3}{l}{$\textit{}$}\\
\multicolumn{18}{l}{$\textit{\{-\# language DuplicateRecordFields      \#-\}}$} & \multicolumn{3}{l}{$\textit{}$}\\
\multicolumn{18}{l}{$\textit{\{-\# language FlexibleInstances          \#-\}}$} & \multicolumn{3}{l}{$\textit{}$}\\
\multicolumn{18}{l}{$\textit{\{-\# language GeneralizedNewtypeDeriving \#-\}}$} & \multicolumn{3}{l}{$\textit{}$}\\
\multicolumn{18}{l}{$\textit{\{-\# language MultiParamTypeClasses      \#-\}}$} & \multicolumn{3}{l}{$\textit{}$}\\
\multicolumn{18}{l}{$\textit{\{-\# language NamedFieldPuns             \#-\}}$} & \multicolumn{3}{l}{$\textit{}$}\\
\multicolumn{18}{l}{$\textit{\{-\# language PartialTypeSignatures      \#-\}}$} & \multicolumn{3}{l}{$\textit{}$}\\
\end{tabular}

%% Acknowledgments
\begin{acks}                            %% acks environment is optional
                                        %% contents suppressed with 'anonymous'
The author thanks for all
tap-on-the-back donations to his past
projects.

We wrote the article with the great help
of bidirectional literate programming
{[}\protect\hyperlink{ref-literate-programming}{15}{]}
tool
{[}\protect\hyperlink{ref-entangled}{7}{]},
Pandoc
{[}\protect\hyperlink{ref-pandoc}{19}{]}
markdown publishing system and live
feedback from GHCid
{[}\protect\hyperlink{ref-ghcid}{12}{]}.
\end{acks}

\bibliography{towards-better-union.bib}

\end{document}